# Evaluation of approaches for accommodating interactions and non-linear terms in multiple imputation of incomplete three-level data

**Running head: Imputing three-level data with interactions and non-linear terms**

Rushani Wijesuriya[1, 2]*, Margarita Moreno-Betancur[1, 2], John B. Carlin[1, 2, 3], Anurika P. De Silva[3] and Katherine J. Lee[1, 2]

*Correspondence:

Email: rushani.wijesuriya@mcri.edu.au

Mobile: +61435302097

[1] Department of Paediatrics, Faculty of Medicine Dentistry and Health Sciences, The University of Melbourne, 50 Flemington road, Parkville, Victoria 3052.

[2] Clinical Epidemiology and Biostatistics Unit, Murdoch Children's Research Institute, 50 Flemington road, Parkville, Victoria 3052.

[3] Centre for Epidemiology and Biostatistics, Melbourne School of Population and Global Health, University of Melbourne, Melbourne, Australia.

## Abstract

Three-level data structures arising from repeated measures on individuals clustered within larger units are common in health research studies. Missing data are prominent in such studies and are often handled via multiple imputation (MI). Although several MI approaches can be used to account for the three-level structure, including adaptations to single- and two-level approaches, when the substantive analysis model includes interactions or quadratic effects these too need to be accommodated in the imputation model. In such analyses, substantive model compatible (SMC) MI has shown great promise in the context of single-level data. While there have been recent developments in multilevel SMC MI, to date only one approach that explicitly handles incomplete three-level data is available. Alternatively, researchers can use pragmatic adaptations to single- and two-level MI approaches, or two-level SMC-MI approaches. We describe the available approaches and evaluate them via simulation in the context of a three three-level random effects analysis models involving an interaction between the incomplete time-varying exposure and time, an interaction between the time-varying exposure and an incomplete time-fixed confounder, or a quadratic effect of the exposure. Results showed that all approaches considered performed well in terms of bias and precision when the target analysis involved an interaction with time, but the three-level SMC MI approach performed best when the target analysis involved an interaction between the time-varying exposure and an incomplete time-fixed confounder, or a quadratic effect of the exposure. We illustrate the methods using data from the Childhood to Adolescence Transition Study.

## 1. Introduction

Three-level data structures are common in observational data. For example in life-course epidemiological studies, clustering of repeated measures from individuals (i.e. longitudinal data) who are further grouped within larger higher-level clusters, or where there are 2 multiple layers of clustering such as participants within classes within schools. The Childhood to Adolescence Transition Study (CATS) is one such study where a cohort of students recruited from schools in Victoria, Australia, is followed up at multiple waves on a range of measures [1]. Missing data are a common challenge in many studies, but can be particularly problematic in longitudinal studies such as the CATS where there are multiple waves of data collection.

Multiple imputation (MI) is a widely used approach for handling missing data [2]. MI is a procedure whereby missing values are imputed multiple times using an imputation model based on the observed data, followed by analysing each imputed dataset using the intended substantive analysis model, and then appropriately pooling the resulting inferences across the multiply imputed datasets [3]. MI approaches impute the missing values from the conditional distribution of missing data given the observed data in an appropriate model for the observed and missing values. There are two common parametric model-based approaches to MI for imputing missing data in multiple variables: (i) joint modelling (JM), which imputes all partially observed variables simultaneously by specifying a joint model for the partially observed variables, conditional on any fully observed variables [4], and (ii) fully conditional specification (FCS, also known as multiple imputation via chained equations) which imputes by iteratively sampling from a series of univariate conditional models for each partially observed variable conditional on all the other variables [5,6]. It has been shown that for MI to generate valid estimates of the substantive model parameters, the imputation model needs to be congenial with the analysis model, that is, it should preserve all the key features of the analysis such as non-linear relationships, interactions and multilevel features [7,8]. The standard JM and FCS methods (referred to as single-level JM and single-level FCS) impute the missing values assuming that the individual observations are independent. However, when the substantive analysis of interest is a multilevel analysis, for example using a linear mixed model (LMM), failing to incorporate the multilevel structure in the imputation models can lead to biased estimates of the regression coefficients and their standard errors (SE), and also may severely bias estimates of the variance components due to the non-congeniality with the analysis model. This is especially so when the proportion of cases with missing data is large [9-12].

There are several approaches that can be used to account for the multilevel structure during the imputation process. A simple way of extending the single-level imputation approaches for imputing incomplete two-level data is to include a series of dummy indicators (DIs) to represent the clusters, although this may not be a viable approach for longitudinal data due to the number of indicators required. For longitudinal data, with follow-ups at fixed intervals of time, a common approach is to arrange the repeated measures of the same variable in wide format and treat each as a distinct variable in the imputation model [13]. Alternatively, JM and FCS MI approaches that are based on multilevel models have recently been developed for handling incomplete multilevel data, with either 2 or 3 levels. Similarly to its single-level counterpart, the multilevel JM approach imputes the missing values simultaneously by specifying a multivariate linear mixed model (MLMM) as the imputation model [14], while the multilevel FCS approach imputes missing values using a series of univariate LMMs [12]. In the context of three-level data such as the CATS, there are a number of approaches available: imputation based on three-level models or extensions of single-level or two-level imputation approaches, using DIs to represent the higher-level cluster membership and/or by including repeated measures as distinct variables (arranged in "wide format"). Although the DI extension for handling multilevel data has been shown to produce biased estimates of substantive model parameters in some cases [11,15,16], in our previous work we found that these pragmatic adaptations to the single-level and two-level MI methods, and the three-level MI methods, all produced valid inferences in the context of a multilevel, random intercepts (at both levels), analysis model [17]. However, similar comparisons of these approaches under more complex substantive analysis models such as when the analysis model includes interactions or quadratic effects involving incomplete covariates are lacking[18].

Several ad hoc ways of handling the interaction and/or quadratic effects using the standard MI approaches have been proposed and evaluated in the literature, although the majority of studies have focused on single-level models [19-21]. One strategy is to "impute-then-transform" also known as "passive imputation", where the non-linear or interaction terms are ignored in the imputation process and passively derived post-imputation. Alternatively, within the FCS framework, "passive imputation" can be incorporated at each iteration, so that the interactions or non-linear terms are updated to incorporate the most recent imputations and can be used to impute the other incomplete variables [22]. Another approach is to "transform-then-impute" the non-linear or interaction term, where the non-linear or interaction term is created prior to imputation and treated as "just another variable" (JAV) in the imputation model [9,19,20,23]. All of these approaches have been shown to produce biased regression coefficient estimates when the substantive model includes interaction or non-linear effects when data are missing at random (MAR) [18,20,24,25]. It has also been suggested to use a reverse imputation strategy, where if the analysis model includes an interaction, the imputation model for the incomplete variable(s) includes an interaction between the outcome and the variables involved in the interaction [25,26]. However, analytical derivations of the conditional distributions of the incomplete covariates for several analysis models involving interactions and non-linear terms clearly show that the conditional distributions for the incomplete covariates specified by all of the above ad-hoc approaches are not compatible with those implied by the substantive analysis model [27-29].

Bartlett et al. (2015) introduced substantive model compatible (SMC) MI which involves generating imputed values from appropriately specified conditional distributions of incomplete covariates for a given substantive model [9,28,30,31]. This is achieved by specifying the imputation model using a decomposition of the joint distribution of all variables into a conditional distribution for the outcome given covariates, one which is set to align with the substantive analysis model, and a joint distribution for the covariates [29,30]. Such an imputation model should generate appropriate imputations for the incomplete variables, as the imputation model and the substantive model are conditional distributions from a common joint distribution [28]. The imputed values from this model can be drawn by using a Markov Chain Monte Carlo method, such as a Gibbs sampler, with an additional Metropolis-Hastings step for imputing missing values for the covariates which will accept or reject the draws to selectively sample from the corresponding substantive model as implied by the imputation model. SMC-MI can be implemented using joint modelling (SMC-JM), fully conditional specification (SMC-FCS) or sequential modelling (SMC-SM), a variation of the JM approach which factorizes the joint distribution of the variables into a sequence of conditional distributions [29]. In the context of single-level data, all these approaches have been shown to perform better than ad-hoc extensions such as JAV and the different passive imputation approaches when the analysis model includes an interaction or a non-linear term [27-29,32]. The single-level implementations of SMC-MI can be implemented in the 'smcfcs' package and command in R and Stata, respectively [33]. While SMC-MI has been extended to the context of multilevel data, the extensions and their evaluations are limited to two-level data with no guidance for the settings of three-level data [34,35]. These extensions have been implemented in REALCOM and Stat-JR software and more recently in the R packages 'jomo','mdmb' and 'JointAI' [36-38]. We are aware of only one SMC-MI approach that has been specifically extended for handling incomplete three-level data, which is implemented in the Blimp software [39] although this approach has not been fully evaluated and has not been compared with pragmatic adaptations such as JAV and passive imputation in single- and two-level MI approaches, or two-level SMC-MI approaches in the context of interactions or non-linear terms with three-level data. Therefore the aim of the current paper is to compare the available approaches for imputing three-level data in the context of substantive analysis models involving interactions or non-linear terms commonly used in longitudinal data settings, using both simulations and a case study. We focus on multilevel data resulting from repeated measures with follow-ups at fixed intervals of time within an individual where there is clustering among individuals as in the CATS.

The remainder of the paper is structured as follows. Section 2 describes the case study that motivated this research and the substantive analysis models of interest. Section 3 provides an overview of the approaches that can be used to impute incomplete three-level data and how they can be used to accommodate the interactions or quadratic effects of covariates. In section 4 we describe a simulation study based on the CATS case study in which we evaluate the available approaches, comparing their performance in terms of bias and precision. In section 5 we apply these approaches to the analysis of the case study. We conclude with a general discussion in section 6.

## 2. The motivating example: The Childhood to Adolescence Transition Study (CATS)

The motivation for this study came from the CATS, a longitudinal cohort study which focuses on educational, emotional, social and behavioural changes in children as they transition from puberty to adolescence [1]. The participants were recruited from a stratified random sample of 43 schools in Melbourne, Australia. All grade 3 students (8-9 years of age) enrolled in these schools were invited to participate. Of the 2239 invited students, 1239 (54%) children with informed parental/guardian consent were recruited into the study at wave 1 (2012). Data were collected through annual follow-ups using parent, teacher and student self-report questionnaires, with currently 7 waves of follow-up available for analysis. There is also linkage with the Victorian Curriculum and Assessment Authority (VCAA) to obtain National Assessment Programme – Literacy and Numeracy (NAPLAN) results. NAPLAN is a nationwide test administered to students in grades 3,5,7 and 9 (approximate ages 8-9, 10-11, 12-13, 14-15 years), which assesses the student's academic performance on 4 domains: reading, writing, numeracy and language conventions. The detailed study protocol can be found elsewhere [1].

### 2.1 Substantive analysis models

The motivating example for our study aimed to estimate the effect of early depressive symptoms (at waves 2, 4 and 6) on the academic performance of the students at the subsequent wave (waves 3, 5, and 7) as measured by NAPLAN numeracy scores with adjustment for potential confounders measured at baseline (wave 1): child's NAPLAN numeracy scores, sex, socio-economic status (SES), and age [40]. With complete data, the standard modelling framework for estimating these effects would be a LMM. We focus on three LMMs involving interactions and quadratic terms, as defined by equations (1), (2) and (3), which are typical of the models that are of interest to researchers in longitudinal data settings. In these model specifications $i$ denotes the school ($i = 1, \ldots, 43$), $j$ denotes the individual ($j = 1, \ldots 1239$) and $k$ denotes the wave of data collection ($k = 3,5,7$), with $\varepsilon_{ijk}$ a random error distributed as $N(0, \sigma_1^2)$ and school and individual-level random effects $\alpha_{0i} \sim N(0, \sigma_3^2)$ and $\alpha_{0ij} \sim N(0, \sigma_2^2)$, respectively.

A detailed description of the variables in these models is provided in <Table 1 on page 21>.

The analysis models of interest are:

1. A random intercept model with an interaction between the time-varying exposure and time

$$NAPLAN_{z_{ijk}} = \beta_0 + \beta_1 * depression_{ij(k-1)} + \beta_2 * wave_{ijk} + \\ \beta_3 * depression_{ij(k-1)} * wave_{ijk} + \beta_4 * NAPLAN_{z_{ij1}} + \\ \beta_5 * sex_{ij} + \beta_6 * SES_{ij1} + \beta_7 * age_{ij1} + \alpha_{0i} + \alpha_{0ij} + \varepsilon_{ijk}$$ (1)

Such a model allows the association between the exposure and the outcome to vary with time.

2. A random intercept model with an interaction between the time-varying exposure and a time-fixed baseline variable

$$NAPLAN_{z_{ijk}} = \beta_0 + \beta_1 * depression_{ij(k-1)} + \beta_2 * wave_{ijk} + \\ \beta_3 * depression_{ij(k-1)} * SES_{ij1} + \beta_4 * NAPLAN_{z_{ij1}} + \\ \beta_5 * sex_{ij} + \beta_6 * SES_{ij1} + \beta_7 * age_{ij1} + \alpha_{0i} + \alpha_{0ij} + \varepsilon_{ijk}$$ (2)

This model allows the association between the exposure and the outcome to vary across strata of the level 2 variable, in this case SES.

3. A random intercept model with a quadratic effect of the time-varying exposure

$$NAPLAN_{z_{ijk}} = \beta_0 + \beta_1 * depression_{ij(k-1)} + \beta_2 * wave_{ijk} + \\ \beta_3 * depression_{ij(k-1)}^2 + \beta_4 * NAPLAN_{z_{ij1}} + \beta_5 * sex_{ij} + \\ \beta_6 * SES_{ij1} + \beta_7 * age_{ij1} + \alpha_{0i} + \alpha_{0ij} + \varepsilon_{ijk}$$ (3)

This model implies that the association between the exposure and the outcome is non-linear.

In the CATS, data were missing for the (time-varying) exposure, depressive symptom score, and the outcome, NAPLAN numeracy scores. NAPLAN numeracy scores were missing for 15% (184/1239) of individuals at wave 1, 16% (198/1239) at wave 3, 21% (264/1239) at wave 5, and 30% (366/1239) at wave 7. Depressive symptom scores were missing for 11% (137/1239) of individuals at wave 2, 14% (173/1239) at wave 4 and 21% (249/1239) at wave 6.

### 3. Overview of possible MI approaches for handling incomplete three-level data in the context of interactions and/or quadratic terms

In this section, we outline the MI approaches that can be used to impute incomplete three-level data in the context of the analysis models outlined in section 2, describing how each approach handles the non-linear or interaction term as well as the two sources of clustering: in our context the correlation among individuals belonging to the same school and the correlation among the repeated measures of an individual.

> ### *i)* *Single-level JM with DI indicators for the higher level clusters and repeated measures imputed in wide format (JM-1L-DI-wide)*

Single-level JM, popularized by Schafer (1997), imputes the missing values by assuming a joint distribution for the incomplete variables. Imputations for the missing values are drawn from the joint posterior predictive distribution of the missing data given the observed data [4]. Commonly the joint distribution is assumed to be multivariate normal (MVN) (including for categorical variables), where the incomplete variables are included as outcomes and the complete variables are included as predictors in the imputation model. This approach can be implemented in most statistical software. A slight variation to the standard MVN approach is where incomplete categorical variables are modelled directly as a part of the joint model by modelling them as normally distributed latent variables[41]. This approach has been implemented in the R package 'jomo' [36].

A simple way of adapting the JM approach to handle three-level data with repeated measures and clustering is to include a set of DIs (a total of $I-1$ DIs for $I$ higher-level clusters) representing the cluster membership of each individual. This estimates a separate intercept/fixed effect for each cluster [11]. The repeated measures can then be analysed in wide format (with one row per individual and separate variables for each repeated measure) to allow for the clustering of repeated measures within an individual. This approach models the clustering of repeated measures in the imputation model by allowing for an unstructured pattern of correlations between them. However, this approach can only be used for repeated measures with follow-ups at fixed intervals of time within an individual, as in CATS.

Under this approach, because the repeated measures are imputed in wide format, the imputation model allows the relationship between the exposure and the outcome to be different at different time points, hence naturally incorporates an interaction between the time-varying exposure and time. Because this approach is congenial with an analysis model with an interaction between a time-varying covariate and time we expect this approach to produce valid results for analysis model 1. In contrast, this approach does not implicitly accommodate interactions between the time-varying exposure and a time-fixed baseline variable (as in analysis model 2) or a non-linear term (as in analysis model 3). To handle these analyses, additional terms can be incorporated into the imputation model using JAV or passively derived after imputation (i.e. the terms can be derived post-imputation or at each iteration respectively, as described in the introduction), although we note that in this case the imputation model is not congenial with the substantive analysis model.

> ### *ii)* *Single-level FCS with DI indicators for the higher level clusters and repeated measures imputed in wide format (FCS-1L-DI-wide)*

An analogous approach to ***JM-1L-DI-wide*** is to use a single-level FCS approach with DI indicators representing the cluster membership of each individual, and imputing the repeated measures in the wide format. In contrast to JM, under the FCS approaches the imputations for the missing values in each variable in turn are drawn using an iterative algorithm that cycles through univariate imputation models [6]. Similarly to ***JM-1L-DI-wide***, including DIs as predictors in these univariate models to model the correlation among individuals belonging to the same higher-level cluster. Meanwhile the imputation model effectively also allows an unstructured correlation matrix between the repeated measures because, when imputing an incomplete repeated measure at one time point/wave, repeated measures at all the other waves are used as predictors.

Similarly to *JM-1L-DI-wide,* this approach naturally accommodates an interaction between the time-varying exposure and time (analysis model 1), but not an interaction with a time-fixed covariate (analysis model 2) or a quadratic term (analysis model 3), which could be incorporated using JAV or one of the variations of passive imputation.

### iii) Two-level JM for the higher level clusters with repeated measures imputed in wide format (JM-2L-wide)

First introduced by Schafer and Yucel (2002), the two-level JM approach is an extension of the single-level JM approach that imputes data using a joint MLMM [14]. The *JM-2L-wide* approach consists of using a two-level JM to model the correlation among individuals within a higher-level cluster using cluster-specific random effects (which are assumed to follow a normal distribution), with the repeated measures within individuals incorporated by imputing the data in wide format as in the previous approaches. Similarly to *JM-1L-DI-wide* and *FCS-1L-DI-wide*, this approach naturally accommodates the interaction between the exposure and time in analysis model 1, with the interaction with a time-fixed covariate (analysis model 2) and the quadratic term (analysis model 3) incorporated using JAV or passively derived post-imputation.

### iv) Two-level FCS for the higher level clusters with repeated measures imputed in wide format (FCS-2L-wide)

Van Buuren (2011) proposed an extension of FCS for imputing two-level data which uses a series of univariate two-level LMMs to impute the missing values, cycling through the incomplete variables one at a time [12]. The *FCS-2L-wide* approach uses a univariate two-level LMM for each incomplete repeated measure with cluster-specific random effects to account for the correlation among individuals of the same higher-level cluster, while repeated measures are imputed as distinct variables in wide format. Again this approach naturally accommodates the interaction in analysis model 1, with the interaction with a time-fixed covariate (analysis model 2) and the quadratic term (analysis model 3) incorporated using JAV or a variant of passive imputation.

### v) Two-level substantive-model-compatible JM for the repeated measures with DI for the higher level clusters (SMC-JM-2L-DI)

Goldstein et al. (2014) proposed a SMC-JM approach for imputing multilevel data, where the imputation model is defined as the product of the substantive model and the joint distribution of the covariates. With this approach, the imputations for incomplete variables are drawn simultaneously using an iterative Gibbs sampler algorithm with a Metropolis–Hastings step [31]. Similarly to *JM-1L-DI-wide* and *JM-2L-wide*, the joint distribution is often assumed to be a MVN distribution, with categorical covariates imputed by assuming underlying latent continuous variables [9,31].

Current implementations of this approach can only handle up to two levels i.e. one type of clustering [36,42]. Therefore in order to impute incomplete three-level data, when conducting the imputation step the substantive model must be specified as a two-level (rather than three-level) LMM with a random effect modelling the repeated measures within an individual, and using DIs to represent the cluster membership. Here the two level approach is used to handle the clustering of repeated measures within an individual as the substantive analysis model requires the time varying exposure to be in long format to model the linear trend with time which cannot be modelled with the data in the wide format. The correct (three-level) substantive model is fitted at the analysis stage to obtain the final parameter estimates.

Under this approach, because the imputation model contains (a modified version of) the substantive model as a corresponding conditional, the interactions and the quadratic effects in our substantive analysis models are naturally incorporated in the imputation model.

### vi) *Two-level substantive-model-compatible sequential modelling for the repeated measures with DI for the higher-level clusters (SMC-SM-2L-DI)*

Ibrahim et al. (2002) proposed an alternative SMC-SM approach where again the imputation model is decomposed into a product of the substantive model and the joint distribution of the covariates, but where the latter is further decomposed into a sequence of univariate conditional models for each incomplete covariate conditioning on the remaining variables [43]. Under this approach, each univariate conditional model can be specified as a generalized linear mixed model (GLMM) or a generalized linear model (GLM) according to the scale of measurement of the incomplete covariate. For example, a probit regression model can be used to impute binary variables and an ordered probit regression model for imputing ordinal variables [29].

Current implementations of this approach are limited to two-level incomplete data. Therefore, similarly to **SMC-SM-2L-DI**, this method can be used to handle incomplete three-level data by setting the substantive model to be a two-level GLMM in the imputation step, where the clustering of repeated measures within an individual is modelled via subject-specific effects (in long format), while the correlation among individuals of the same higher-level cluster is modelled by DIs. Each incomplete covariate can then be modelled using an appropriate univariate conditional model. For incomplete time-varying (level 1) covariates this would be a GLMM with subject-specific effects to model the within-individual correlation and DIs to model the higher-level clustering, while with time-fixed (level 2) covariates a GLM with just the DIs to model the higher-level clustering suffices. For modelling covariates measured at the higher-level clusters (level 3), an appropriate GLM can be used. The correct substantive model is then fitted in the analysis step.

Similarly to **SMC-JM-2L-DI**, because the imputation model specified contains (a modified version of) the substantive model as a nested conditional distribution, the interactions and the quadratic effect in the substantive analysis models are naturally accommodated in the imputation model.

### vii) *Three-level substantive-model-compatible JM (SMC-JM-3L)*

Finally, Enders at al. (2019) have developed a three-level SMC approach, which is implemented in the Blimp software [39]. Similarly to **SMC-JM-2L-DI**, the imputation model is decomposed into the product of the substantive model and the joint distribution of the covariates, but in this case the substantive model is defined as a three-level LMM. Within this approach the correlation among individuals within the same higher-level cluster is modelled using cluster-specific random effects and the correlation among repeated measures within individuals via subject-specific effects (in long format) [28]. As with **SMC-JM-2L-DI,** the joint covariate distribution assumed is a MVN distribution, with categorical covariates handled by assuming underlying latent continuous variables. However, in contrast to **SMC-JM-2L-DI**, this approach imputes the incomplete covariates one at a time by factorising the implied joint distribution of the covariates into the univariate conditional distribution of each incomplete covariate using Gibbs sampling with a final Metropolis–Hastings step. Incomplete time-varying (level 1) variables are imputed using a three-level random intercept model, regressing on all the other covariates. Time-fixed (level 2) variables are imputed using a two-level random intercept model regressing on all the other level 2 and level 3 variables, and the cluster means of the level 1 variables. Similarly, higher-level cluster specific variables (level 3) are imputed using linear regression on all the level 3 covariates and the level 1 and level 2 cluster means. Of note, the cluster means can either be calculated as the arithmetic averages of the variable at the required level (known as the "manifest" approach) or can be modelled as normally distributed latent variables (known as the "latent" approach) [35].

Similarly to **SMC-JM-2L-DI** and **SMC-SM-2L-DI**, because the imputation model is specified in a way that includes the substantive model as a conditional, the interactions and the quadratic effect in the substantive analysis models are naturally accommodated in the imputation model.

<Table 2 on page 23> provides a summary of all the approaches discussed above along with details of the software packages within which each approach is available.

Note: In this study, we do not consider any SMC-FCS approaches as multilevel extensions of this approach are not currently available and the single-level implementations cannot be carried out with the data in the wide format.

### 4. Simulation Study

*4.1 Simulation of complete and missing data*

We conducted a simulation study to compare the performance of the MI approaches detailed above in the context of the three-level analysis model outlined in equations 1-3. Data were generated as described below to obtain simulated datasets with sample size 1200 students for each scenario. The number of simulations for each scenario was 1000, chosen to limit the Monte Carlo error for the coverage of nominal 95% confidence intervals to approximately 0.7% [44].

For each analysis model, we considered scenarios with two different cluster sizes: one scenario with 40 schools (clusters) each with 30 students, similar to CATS, and another with 10 clusters each with 120 students, to provide a scenario with a smaller number of higher-level clusters. The variables were generated sequentially as described below for individual $j$ in cluster $i$. The values of the parameters indexing these distributions were determined by estimating the respective quantity from the CATS data and are given in Additional file 1: Table S1.

i. Individuals were generated within school clusters under the 2 scenarios described above.

ii. Child's age at wave 1 ($age_{ij1}$) was generated from a uniform distribution, $U(a,b)$.

iii. Child's sex ($sex_{ij}$) was generated by randomly assigning $\lambda\%$ of students to be female.

iv. Child's standardized SES value at wave 1 ($SES\_z_{ij1}$) was generated from a standard normal distribution, $N(0,1)$

v. The standardized NAPLAN scores at wave 1 ($NAPLAN\_z_{ij1}$) were generated from a linear regression model conditional on child's sex, child's age at wave 1 and child's SES value:

$$NAPLAN\_z_{ij1} = \eta_0 + \eta_1 * [sex_{ij} = 1] + \eta_2 * age_{ij1} + \eta_3 * SES\_z_{ij1} + \psi_{ij} \qquad (4)$$

where $\psi_{ij}$ are independently and identically (iid) distributed as as $\psi_{ij} \sim N(0, \sigma_\varphi^2)$

vi. To generate the time-varying variables, each student record was expanded to include five repeated observations (from waves 2-6)

vii. Child's depression status at waves 2, 4 and 6 ($depression_{ij(k-1)}$) was generated using a LMM conditional on child's age at wave 1, child's sex, NAPLAN scores at wave 1, child's SES value and wave:

$$depression_{ijk} = \delta_0 + \delta_1 * age_{ij1} + \delta_2 * [sex_{ij} = 1] + \delta_3 * NAPLAN\_z_{ij1} + \delta_4 * SES\_z_{ij1} + \delta_5 * wave_{ijk} + u_{0i} + u_{0ij} + \varphi_{ijk} \qquad (5)$$

where $\varphi_{ijk}$, $u_{0ij}$ and $u_{0i}$ are iid as $\varphi_{ijk} \sim N(0, \sigma_\varphi^2)$, $u_{0ij} \sim N(0, \sigma_{u_2}^2)$, and $u_{0i} \sim N(0, \sigma_{u_3}^2)$ respectively.

viii. The outcome, child's standardized NAPLAN score at waves 3, 5 and 7 ($NAPLAN\_z_{ijk}$), was generated using the relevant target analysis model as per equations (1), (2) and (3).

ix. Finally, we generated child's behavioural problems at waves 2, 4 and 6 ($SDQ_{ijk}$), which is not included in the analysis model but is associated with the exposure, was generated using a LMM conditional on depression symptoms at waves 2, 4 and 6 and wave:

$$SDQ_{ijk} = \gamma_0 + \gamma_1 * depression_{ijk} + +\gamma_2 * wave_{ijk} + v_{0i} + v_{0ij} + \epsilon_{ijk} \qquad (6)$$

where $\epsilon_{ijk}$, $v_{0i}$ and $v_{0ij}$ are iid as; $\epsilon_{ijk} \sim N(0, \sigma_\epsilon^2)$ $v_{0i} \sim N(0, \sigma_{v_3}^2)$, and $v_{0ij} \sim N(0, \sigma_{v_2}^2)$ respectively. This is an example of an auxiliary variable that can be included in the imputation model to improve its performance [45],

*4.2 Generation of missing data*

To simulate missingness, data were set to missing in depressive symptom scores at waves 2, 4 and 6 (the exposure of interest) and SES at wave 1. In our simulation study, for depression symptom scores, the proportions of missingness were set around 15%, 20% and 30% at waves 2, 4 and 6, respectively which were generated according to two MAR mechanisms, labelled MAR-CATS and MAR-inflated, by drawing from a logistic regression model dependent on the standardized NAPLAN scores at the subsequent wave and the SDQ measure at the concurrent wave:

$$\text{logit}(P(R\_depression_{ijk} = 1)) = \zeta_{0k} + \zeta_1 * NAPLAN\_z_{ij(k+1)} + \zeta_2 * SDQ_{ijk} \quad (7)$$

where, $R\_depression_{ijk}$ is an indicator variable that takes the value 0 if $depression_{ijk}$ is missing and 1 if $depression_{ijk}$ is observed.

For the MAR-CATS scenario we set $\zeta_1$=1.5 and $\zeta_2$ =2 which represent the associations between the probability of response and the predictors of response as observed in CATS. For the MAR-inflated scenario the values of $\zeta_1$ and $\zeta_2$ were doubled. The values of the intercepts $\zeta_{0k}$ were chosen by iteration so that the required proportions of missingness were achieved for each of the waves (2, 4 and 6). For simplicity we set around 10% of the individuals to have missing SES values according to a missing completely at random (MCAR) mechanism using simple random sampling.

*4.3 MI methods and evaluation*

For each of the 12 scenarios considered (3 substantive analysis models x 2 cluster sizes x 2 missingness mechanisms), we applied the 7 MI approaches (***JM-1L-DI-wide***, ***FCS-1L-DI-wide, JM-2L-wide***, ***FCS-2L-wide, SMC-JM-2L-DI, SMC-SM-2L-DI*** and ***SMC-JM-3L***), using ad hoc extensions to handle the interactions or non-linear terms where necessary, to impute missing values in depressive symptom scores at waves 2, 4 and 6 and SES at wave 1 in each of the simulated data sets (as described below).

Specifically, for analysis model 1, which involves an interaction between the depressive symptom scores and time, we applied each of the 7 MI approaches to impute the missing values in SES at wave 1 and the depressive symptom scores at waves 2, 4 and 6. No variations of these approaches were considered as all 7 approaches are congenial with the analysis model.

For analysis model 2, which involves an interaction between the depressive symptom scores and SES, ***JM-1L-DI-wide*** and ***JM-2L-wide*** were applied using JAV to incorporate the interaction by including the interactions between depressive symptom scores at each wave (2, 4 and 6) and SES as distinct variables in the imputation model (denoted as ***JM-1L-DI-wide-JAV*** and ***JM-2L-wide-JAV***). The JAV approach was used here as it has been shown in previous simulation studies to perform better than passive imputation post-imputation which is the only passive approach possible with JM [19,20]. ***FCS-1L-DI-wide*** and ***FCS-2l-wide*** were implemented using passive imputation carried out within iterations using two variations of reverse imputation strategy[18,25]:

i) In the first approach (denoted ***FCS-1L-DI-wide-passive_c*** and ***FCS-2L-wide-passive_c***), we included the single interaction between the NAPLAN score at the next wave and SES as a predictor in the imputation model for depressive symptom scores and the two-way interactions between the NAPLAN scores and depressive symptom scores at previous wave for all 3 waves as predictors when imputing SES. This method allows the association between the outcome and exposure at each wave to vary for different levels of SES and vice versa as implied by the substantive analysis model.

ii) In the second approach (denoted *FCS-1L-DI-wide-passive_all, FCS-2L-wide-passive_all*), we included the two-way interactions between the NAPLAN scores at each of the 3 waves and SES as predictors in the imputation model for depressive symptoms at each wave and the two-way interactions between the NAPLAN scores and depressive symptom scores at previous wave for all 3 waves as predictors when imputing SES. Similar to the first approach, this method allows the association between the outcome and the exposure to vary for different levels of SES and vice versa, but allows even more flexibility. (*Note:* although it is possible to include all the two-way interactions between NAPLAN scores and depressive symptom across the waves can be included as predictors when imputing SES allowing further flexibility, we opted to only include the ones implied by the analysis model for simplicity).

For this analysis model we also present the results from the *JM-1L-DI-wide* and *FCS-1L-DI-wide* approaches where we do not incorporate the interaction as a benchmark for comparison, resulting in a total of 11 approaches for comparison for this analysis model.

For analysis model 3, which involves a quadratic effect of the exposure, for *JM-1L-DI-wide* and *JM-2L-wide* we use a JAV approach for the quadratic term of the exposure, while for *FCS-1L-DI-wide* and *FCS-2l-wide* we use passive imputation of the quadratic term within iterations. Again we also present results from the *JM-1L-DI-wide* and *FCS-1L-DI-wide* approaches where we do not incorporate the quadratic term as a benchmark for comparison, resulting in a total of 9 approaches for comparison for this analysis model

Except for *SMC-JM-3L* in Blimp, all approaches were implemented in R version 3.6.1 [46]. *JM-1L-DI-wide*, *JM-2L-wide* and *SMC-JM-2L-DI* were implemented in the R package 'jomo' using the functions jomo1con, jomo1rancon and jomo.lmer respectively, while *FCS-1L-DI-wide* and *FCS-2L-wide* were implemented in the R package 'mice' using the functions norm and 2l.pan respectively [36,47]. While there are several alternative functions in R for specifying a two-level FCS imputation model, we chose the above because they are the most established. The sequential modelling approach, *SMC-SM-2L-DI,* was implemented using the function frm_fb in the R package `mdmb'[37]. For *SMC-JM-3L* we used the Blimp version 1.0.6 with the default "latent" specification [39].

In addition to all the variables in the analysis model, the imputation model for each of the MI approaches also included child behaviour problems (measured by SDQ) at waves 2, 4 and 6 as auxiliary variables. For each simulated dataset, 20 imputations were generated for each of the MI approaches [9]. After examining trace plots, the JM and SMC-SM approaches in R were run with a burn-in of 1000 iterations and 100 between-imputation iterations, while the SMC-JM approach in R were run with a burn-in of 500 iterations and 10 between-imputation iterations. The FCS approaches were run with a burn-in of 10 iterations. The SMC-JM approach in Blimp was run with a burn-in of 2500 iterations with 100 thinning iterations, as with 2500 iterations the highest (worst) potential scale reduction (PSR) factor value across all parameters was generally less than 1.10 [48]. These values were further confirmed by examining trace plots.

The substantive analysis models were fitted to each of the imputed data sets and the resulting parameter estimates pooled using Rubin's rules [49,50]. For each analysis model, the parameters of interest were the regression coefficients corresponding to the main effect of the exposure and the interaction or quadratic term, and the variance components at levels 1, 2 and 3. The estimates of these parameters obtained from the MI approaches were compared to the true values that were used to simulate the data. In order to compare the performance of the various approaches for estimating the regression coefficients of interest we calculated, across 1000 replications, the mean value of the estimate, the bias (the average difference between the true value and the estimates), the empirical standard error (the average standard deviation of the estimates), the model-based standard error (the average of the estimated standard errors of the estimates), and the coverage probability of the nominal 95% confidence interval (estimated as the proportion of replications in which the estimated interval contained the true value). [51] For the variance component estimates, we report the bias and empirical standard error. We also report the percentage bias defined as the bias relative to the true value.

*4.4 Simulation Results*

As the comparative performance of the MI approaches was quite similar for the MAR-CATS and MAR-inflated scenarios we largely focus on the results from the MAR-CATS scenario highlighting contrasts where they exist.

Analysis model 1: Interaction between the time-varying exposure and time

The sampling distribution of the estimated bias of the regression coefficients of interest for analysis model 1 (the main effect $\beta_1$ and the interaction effect $\beta_3$) across the 1000 replications for each MI approach, for the two different higher-level cluster settings, is displayed in Figure 1. While all the MI approaches produced approximately unbiased estimates of these parameters, slightly higher biases were observed for ***SMC-SM-2L-DI*** and ***SMC-JM-3L***.

All the MI approaches also resulted in appropriate nominal coverage and comparable empirical and model-based standard errors for both of the regression coefficients across the two simulation scenarios (Figure 2 and Table S2). However, with a larger number of higher-level clusters ***SMC-JM-2L-DI*** resulted in slightly higher empirical compared with model-based standard errors.

Figure 3 shows the sampling distribution of the estimated biases for the variance components at level 1, 2 and 3 across the two simulation scenarios. All approaches resulted in similar negligible bias (<10% relative bias) for the 3 variance components for both scenarios with slightly larger biases for the level 3 variance estimates when there was a smaller number of higher-level clusters.

Analysis model 2: Interaction between the time-varying exposure and a time-fixed baseline variable

For analysis model 2 the estimates of the main effect were approximately unbiased across all MI approaches, but there were substantial differences in the bias for estimating the interaction effect (Figure 4). As expected, ***JM-1L-DI-wide*** and ***FCS-1L-DI-wide*** (which do not incorporate the interaction in the imputation model) resulted in the largest bias for the interaction term, attenuating it towards zero (Table S6). While including the interaction term as a distinct variable within the JM approaches (*JM-1L-DI-wide-JAV* and *JM-2L-wide-JAV*) and using passive imputation within the FCS approaches (*FCS-1L-DI-wide-passive_c*, *FCS-1L-DI-wide-passive_all, FCS-2L-wide-passive_c* and *FCS-2L-wide-passive_all*) resulted in reduced bias compared to the naïve methods (*JM-1L-DI-wide* and *FCS-1L-DI-wide)*, these approaches produced larger bias than the SMC MI approaches (*SMC-JM-2L-DI, SMC-SM-2L-DI* and *SMC-JM-3L*). The bias in the interaction effect estimates under the FCS approaches that include all the interactions as predictors when imputing the incomplete depressive symptom scores (*FCS-1L-DI-wide-passive_all* and *FCS-2L-wide-passive_all*) were smaller than the FCS approaches that include the single interaction as implied under the analysis model (*FCS-1L-DI-wide-passive_c* and *FCS-2L-wide-passive_c).*

Differences between the model-based and empirical standard errors for the main effect were small for all MI approaches, with somewhat larger discrepancies when the number of higher-level clusters was large under the MAR-inflated scenario (Figure S5). While *JM-1L-DI-wide-JAV, JM-2L-wide-JAV, SMC-SM-2L-DI* and *SMC-JM-3L* provided model-based standard errors that were largely consistent with the empirical standard errors for the interaction effect, *JM-1L-DI-wide-JAV* and *JM-2L-wide-JAV* resulted in upward-biased estimates of the model based standard errors. Somewhat larger discrepancies between the model-based and empirical standard error were observed for the interaction effect for all of the other MI approaches across the simulation scenarios considered (Figure 5, Table S6 and S8).

All of the MI approaches resulted in negligible bias (<10% relative bias) for the variance components at level 1, 2 and 3 across the different simulation scenarios, albeit slightly larger for the level 3 and level 2 variance components when there were fewer higher-level clusters (Figure 6,Table S7).

Analysis model 3: Quadratic term in the exposure

All of the MI approaches except for *SMC-JM-2L-DI, SMC-SM-2L-DI* and *SMC-JM-3L* resulted in biased estimates of the regression coefficients for the main effect and the quadratic term in analysis model 3, with substantial underestimation of the quadratic effect (Figure 7 and Table S10).

Discrepancies between the model-based and empirical standard errors for the quadratic effect were observed for *JM-1L-DI-wide*, *FCS-1L-DI-wide, FCS-1L-DI-wide-passive* and *FCS-2L-wide-passive* (Figure 8). However these approaches retained approximate nominal coverage for the quadratic term under all scenarios. JM approaches using JAV to handle the quadratic effect (*JM-1L-DI-wide-JAV* and *JM-2L-wide-JAV*) resulted in upward-biased model-based standard errors for both coefficient estimates compared to other MI approaches. Under-coverage of the quadratic effect was also observed for these two approaches which was more pronounced under MAR-inflated scenario (Table S12).

Figure 9 shows the estimated bias for the variance components at levels 1, 2 and 3 across different simulation scenarios. All approaches resulted in negligible bias (<10% relative bias) for the variance components across the different simulation scenarios, with slightly larger bias for the level 3 and level 2 variance estimates when there was a smaller number of higher-level clusters (Tables S11 and S13).

## 5. CATS case study illustration

We applied the same MI approaches (7, 11, or 9 depending on the analysis model) as used in the simulations to the three target analysis models in the CATS data. Twenty imputations were generated under each of the MI approaches. Similarly to in the simulation study, the imputation model for each of the MI approaches included all the variables in the analysis model and also the child behaviour problems measures (reported by SDQ) at waves 2, 4 and 6 as auxiliary variables. However in the CATS, missing values occurred in the outcome (NAPLAN scores at waves 3, 5 and 7), the exposure (depressive symptom scores at waves 2, 4 and 6), and a time-fixed baseline variable (NAPLAN scores at wave 1). The auxiliary variable (SDQ at waves 2, 4 and 6) was also incomplete with values missing for 29% (363/1239) of individuals at wave 2, 35% (436/1239) at wave 4 and 24% (297/1239) at wave 6. Tables 3, 4 and 5 < on pages 25 and 26> show the estimated regression coefficients and their standard errors for the main effect and the interaction/quadratic effect, and the variance component estimates for analysis models 1, 2 and 3, respectively, under the different imputation approaches.

Analysis model 1: Interaction between the time-varying exposure and time

As shown in <Table 3 on page 25> , the estimates for the main effect and the corresponding standard errors were very similar for all the MI approaches except for *JM-1L-DI-wide* and *SMC-SM-2L-DI,* which resulted in smaller estimates. Estimates of the interaction effect and standard errors were similar across all approaches, while estimates of the level 3 variance component were smaller for *SMC-JM-2L-DI* and *SMC-SM-2L-DI,* with comparatively larger level 2 variance components for *SMC-JM-2L-DI.*

Analysis model 2: Interaction between the time-varying exposure and a time-fixed baseline variable

<Table 4 on page 25> shows the results from the 11 MI approaches applied to analysis model 2 in the CATS data. The main effect coefficient estimates and their standard errors were very similar irrespective of the analysis method. While the estimates for the interaction effect were similar across most approaches, larger estimates were observed for *JM-1L-DI-wide-JAV* and *JM-2L-wide-JAV.* Similar to analysis model 1, variance component estimates were generally similar although there were smaller estimates for the level 3 variance using *SMC-JM-2L-DI* and *SMC-SM-2L-DI* with slightly larger estimates for the level 2 variance component for *SMC-SM-2L-DI*.

Analysis model 3: Non-linear relationship with the exposure

The results from the MI approaches applied to analysis model 3 in the CATS data is shown in <Table 5 on page 25>. In comparison to the other MI approaches, the regression coefficient estimates for the main effect were larger while the quadratic effect estimates were smaller for *JM-1L-DI-wide, FCS-1L-DI-wide* and *FCS-1L-DI-wide-passive*. Similarly to with analysis models 1 and 2, the variance component estimates were generally similar across the approaches, with smaller estimates for the level 3 variance components with *SMC-JM-2L-DI* and *SMC-SM-2L-DI* and comparatively larger estimates for the level 2 variance component for *SMC-SM-2L-DI.*

**6. Discussion**

While adaptations to single- and two-level MI approaches, using DIs and imputing in wide format, and three-level MI approaches can all be used to accommodate the sources of correlation in three-level data in the imputation process, when the substantive analysis model includes interactions or quadratic effects involving incomplete covariates these too need to be incorporated in the imputation model. While SMC MI has shown great promise in accommodating such terms, there are limited SMC MI implementations that explicitly handle three-level data. In this study we evaluated the performance of several adaptations to currently available MI approaches and the single implementation of SMC MI that was designed to accommodate three-level data, for handling incomplete three-level data in the context of three commonly used LMMs in the analysis of longitudinal data which include an interaction or a non-linear term via simulations and a real data example based on the CATS.

When the analysis model included an interaction between the exposure and time, all of the MI approaches resulted in approximately unbiased estimates of the main effect and the interaction effect, with appropriate coverage, across the different simulation scenarios considered in the simulation study. However, when the analysis models involved an interaction between the exposure and a baseline confounder, or a quadratic effect of the exposure, the approaches which used ad hoc extensions of single- and two-level models resulted in biased estimates of the interaction and the non-linear effects. In contrast, the two-level SMC approaches extended with DIs to handle the second level of clustering, ***SMC-JM-2L-DI*** and ***SMC-SM-2L-DI,*** showed similar performance to the three-level SMC approach, ***SMC-JM-3L***, all of which resulted in approximately unbiased estimates of the interaction or non-linear effects and the variance components. However in the CATS application the former two approaches resulted in comparatively smaller level 3 variance components than the latter. These results suggest that substantive model compatibility is crucial for ensuring appropriate imputations as shown in previous literature in different contexts [27,28,30,31,52].

While there have been recent developments in SMC MI approaches that have shown great promise for accommodating interactions and/or non-linear terms appropriately in the MI literature, implementations of these approaches for multilevel data are limited. In fact, commonly used statistical packages Stata and SPSS do not have SMC-MI approaches, or indeed any MI approaches, which use multilevel imputation models. Therefore, when the substantive analysis model involves an interaction with time, for users of these packages, the only option is to use the single-level MI approaches with pragmatic adaptations, namely ***JM-1L-DI-wide*** and ***FCS-1L-DI-wide***. However, both of these approaches can only be used when repeated measures are recorded at fixed intervals of time. In addition, although we observed approximately unbiased results from these approaches, previous simulation studies have shown the DI approach can result in inflated standard errors and biased variance components estimates (and therefore ICCs) particularly when the ICC is low with a high percentage of missing values and there are small cluster sizes [11,15]. Therefore these approaches should be used with caution. Both these approaches can also be infeasible with a large number of clusters.

Our results confirm, as expected theoretically and from past simulations [28,35], that the three-level SMC approach, ***SMC-JM-3L*** is the most appropriate approach for handling incomplete three-level data where the analysis model includes an interaction among incomplete covariates or quadratic effects. However, this approach is only available in the stand-alone software Blimp [39]. While the two-level SMC MI approaches with DIs used to handle the higher-level clustering, ***SMC-JM-2L-DI*** and ***SMC-SM-2L-DI*** may be potentially useful alternatives, as made evident by the CATS application in our study and previous simulation studies, the DI approach can be problematic when the ICC is low and there is a high percentage of missing values or if there are small cluster sizes [11,15]. We also note that with these two approaches, a modified version of the analysis model must be used in the imputation procedure and the actual substantive analysis model needs to be fitted to the imputed data following imputation. This is in contrast to when these approaches are used to impute two-level data, where the parameters of interest in the substantive analysis model are estimated within the imputation process itself [52]. The **SMC-SM-2L-DI** approach is somewhat more flexible than SMC-JM approaches because each incomplete covariate is modelled separately, which means that non-linear associations among covariates can also be accommodated in the imputation model. This comes at a cost, as this approach requires specification of separate models for each incomplete covariate and ordering of these models which requires more consideration than the specification of the imputation model under SMC-JM approaches [53]. One strategy suggested is to order the incomplete variables by their type and start by conditioning the categorical variables on the continuous variables. Other strategies include ordering the variables so that the missing pattern is close to monotone or according to the percentage of missing data, i.e. starting with those with the least percentage of missing values [29,30,53]. While theoretically the order in which the conditional models for the covariates are specified may result in different joint distributions, the approach has been shown to be robust under different orderings as long as the conditional models for each covariate "fit the data well enough" [52].

In addition to the differences discussed previously, there are also some differences between the two-level JM approach by Schafer and Yucel (2002), as implemented in R, and the two-level FCS implementation in R as originally proposed by van Buuren (2011). The multilevel JM approach allows associations between incomplete variables to vary at different levels, whereas the FCS approach does not allow this [54]. The two approaches can, however, be made equivalent by including the cluster means of the imputed lower level variables in the FCS imputation model. This difference can be important in the context of a multilevel substantive analyses that assumes different associations between variables at different levels. Given the substantive analysis model considered in the current manuscript does not assume such relations, these differences were not relevant here. Further discussion of the differences between the JM and FCS approaches in the multilevel context can be found in Carpenter and Kenward (2013), Enders et al. (2016) and Mistler (2017)[9,54,55].

To our knowledge, there are no previous studies that compare all of the available MI approaches for accommodating interaction and/or non-linear terms in a three-level data setting. However, the results presented in our study are consistent with previously published studies in the single- and two-level settings. Similarly to our findings, several simulation studies in the single-level context have shown that the JAV approach results in biased estimates of the interaction effect and some undercoverage when data are MAR [20,30]. Our results are also consistent with those of Grund (2018), who showed that passive imputation within iterations using the reverse imputation strategy to accommodate interactions in a two-level model resulted in biased estimates of the interaction effect in the presence of incomplete covariates[25]. In contrast, however, Tilling et al (2016) reported approximately unbiased results using a reverse imputation strategy with the JAV approach, their simulations involving an interaction between two binary covariates[26]. Consistent with our results, several studies in the single-level context have also shown the SMC MI approaches perform better than MI approaches with ad hoc extensions such as JAV for accommodating interactions or non-linear effects [27,29,32]. Finally, Enders et al (2019) and Keller (2019) presented a number of simulations evaluating the performance of the *SMC-JM-3L* implementation in Blimp against the JAV approach in a wide range of single-level and multilevel regression models with interaction and non-linear effects, which showed that *SMC-JM-3L* generally resulted in approximately unbiased estimates of the model parameters [28,35].

Designing our simulation study on a real study allowed us to incorporate complex yet realistic associations, hence we believe our results reflect what could be expected in practice in a similar setting. However, we recognize that the simulation conditions we examined are limited in scope. There are a number of factors at play when analysing multilevel data such as the cluster sizes, number of clusters, intra class correlations, effect sizes and the variability of the random effects, and the performance of the methods evaluated in our study may vary with these factors. Therefore caution is required in generalizing these results to conditions outside those evaluated in our study.

Our study is restricted to substantive analyses where the interactions involve incomplete continuous predictors. When the interaction term involves different types of variables, we expect the *SMC-JM-3L* to perform well based on previous simulation studies [35], but the performance of other MI approaches may vary, particularly those that use the DI extension. In the special case when one of the variables involved in the interaction is fully observed and categorical, an alternative approach that can be used, provided that there are not too many categories, is to split the data into strata as defined by this variable and carry out imputation separately within each stratum [26]. Given the limited number of three-level SMC MI implementations, a useful extension of our study would be to evaluate the performance of all the available imputation approaches for incomplete categorical variables in three-level data. Our simulations were also limited to a random-intercept model. Previous research suggest that SMC-MI methods perform well in accommodating random slopes in multilevel models but we expect the performance of the adaptations in this context to be different [28,56]. However, this was beyond the scope of our paper and still an area for future research. Finally, our simulations only considered MAR missingness mechanisms while in practice the data may be missing not at random (MNAR). While the SMC-SM approach may be adapted to handle MNAR data [29], all of the MI approaches considered in our simulations are only guaranteed to produce unbiased estimates under MAR [57]. Therefore examining the performance of these methods under MNAR mechanisms may also be an avenue for future research.

In conclusion, in this study we have shown that the single- and two-level MI approaches, or two-level SMC-MI approaches extended with DIs and/or imputing repeated measures in wide format perform as well as the three-level SMC-MI approaches in accommodating interactions between time-varying exposures and time in the substantive analysis model. Hence in such a context, practitioners may use any of these approaches. However, approaches which use the DI extension should be used with caution as they can be problematic in certain scenarios. We recommend SMC three-level MI approaches be used for handling incomplete three-level data when the substantive analysis includes an interaction between the time-varying exposure and an incomplete time-fixed confounder, or a quadratic effect of the exposure.

**Funding**

This work was supported by funding from the National Health and Medical Research Council: Career Development Fellowship ID#1127984 (KJL) and project grant, ID#APP1166023. Research at the Murdoch Children's Research Institute is supported by the Victorian Government's Operational Infrastructure Support Program. MMB is the recipient of an Australian Research Council Discovery Early Career Award (project number DE190101326) funded by the Australian Government. The funding bodies do not have any roles in the collection, analysis, interpretation and writing the manuscript.

**Data Sharing- Data Availability Statement**

The Childhood to Adolescence Transition Study (CATS) data that support the findings of this study are available on request from the corresponding author. The data are not publicly available due to privacy or ethical restrictions. The software code written for simulating the data that support the findings of this study are openly available in a public GitHub repository at https://github.com/rushwije/MI_three-level.

Table 1: Summary of variables of interest, in the motivating case study in the CATS

| Variable | Type | Grouping /Range | Label |
|---|---|---|---|
| **Child's sex** | Categorical | 0 = Female 1 = Male | $sex_{ij}$ |
| **Child's age (wave 1) (years)** | Continuous | Range [7-11] | $age_{ij1}$ |

| Standardized SES measured by the SEIFA IRSAD (wave 1) | Continuous | z-score | $SES\_z_{ij1}$ |
|---|---|---|---|
| Standardized NAPLAN numeracy score (wave 1) | Continuous | z-score | $NAPLAN\_z_{ij1}$ |
| Standardized NAPLAN numeracy score ($k$ =wave 3,5 and 7) | Continuous | z-score | $NAPLAN\_z_{ijk}$ |
| Depressive symptoms ($k$ =waves 2,4 and 6) [a] | Continuous | Range [0-8] | $depression_{ij(k-1)}$ |
| Overall child behaviour reported by SDQ ($k$ =waves 2,4 and 6) [b] | Continuous | Range [0-40] | $SDQ_{ij(k-1)}$ |

IRSAD: Index of Relative Socio-economic Advantage, NAPLAN: National Assessment Program - Literacy and Numeracy, SDQ: Strengths and Difficulties Questionnaire, SEIFA: Socioeconomic Index for Areas, SES: Socio-Economic Status.

[a] A subset of 4 items (each ranging from 0 to 2) from the Short Mood and Feelings Questionnaire (SMFQ) was used to measure the depressive symptoms at each wave in the CATS study [1,58]. These items are summed and centred at the mean to give a depressive symptoms score.

[b] For measuring the overall child behaviour, a total behavioural difficulties score is derived by summing the first 4 subscales of the Strengths and Difficulties Questionnaire (SDQ): emotional symptoms, conduct problems, hyperactivity/inattention, peer relationship problems (each ranging from 0 to 10) [59]

Table 2: Summary of the imputation approaches for handling incomplete three-level data

| MI approach | Paradigm | Type | Software[*] | How the two sources of clustering are handled | | How the approach accommodate interactions/non-linear terms | | |
|---|---|---|---|---|---|---|---|---|
| | | | | Clustering due to higher level clusters | Clustering due to repeated measures | Interaction between the time-varying exposure and time | Interaction between the time-varying exposure and a time-fixed baseline variable | Quadratic effect of the exposure |
| *JM-1L-DI-wide* | JM | Standard (single-level) | SAS, SPSS, Stata, Mplus, R | DI | Repeated measures imputed in wide format | Repeated measures imputed in wide format | Not accommodated (ad-hoc extensions such as JAV or passive imputation can be used but are not congenial with substantive analysis) | Not accommodated (ad-hoc extensions such as JAV or passive imputation can be used but are not congenial with substantive analysis) |
| *FCS-1L-DI-wide* | FCS | Standard (single-level) | SAS, SPSS, Stata, Mplus, R, Blimp | DI | Repeated measures imputed in wide format | Repeated measures imputed in wide format | | |
| *JM-2L-wide* | JM | Specialized for two levels | SAS, Mplus, Realcom-impute, Stat-JR, R | RE | Repeated measures imputed in wide format | Repeated measures imputed in wide format | | |
| *FCS-2L-wide* | FCS | Specialized for two levels | Mplus, R, Blimp | RE | Repeated measures imputed in wide format | Repeated measures imputed in wide format | | |
| *SMC-JM-2L-DI* | JM | Specialized for two-levels | R, Realcom-impute, Stat-JR | DI | RE | Through SMC-MI algorithm [+] | Through SMC-MI algorithm [+] | Through SMC-MI algorithm [+] |
| *SMC-SM-2L-DI* | SM | Specialized for two-levels | R | DI | RE | Through SMC-MI algorithm [+] | Through SMC-MI algorithm [+] | Through SMC-MI algorithm [+] |

| SMC-JM-3L | JM | Specialized for three-levels | Blimp | RE | RE | Through SMC-MI algorithm [++] | Through SMC-MI algorithm [++] | Through SMC-MI algorithm [++] |

DI: dummy indicators, FCS: fully conditional specification, JM: joint modelling, RE: random effects, SMC: Substantive model compatible, SM: sequential modelling

[*] R and Blimp are the only freely available, open-source software implementations.

[+] The interactions or quadratic effects are accommodated as the imputation model specified under these approaches contains (a 2-level version of) the substantive model as a corresponding conditional.

[++] The interactions and quadratic effects are accommodated as the imputation model specified under this approach contains the substantive model as a corresponding conditional.

Table 3: Point estimate (and standard error) for the main effect and the interaction effect, and point estimates for the variance components at levels 3, 2 and 1, from 7 MI approaches applied to the CATS data analysis in analysis model 1.

| Method | Main Effect (SE) | Interaction Effect (SE) | Level 3 variance component | Level 2 variance component | Level 1 variance component |
|---|---|---|---|---|---|
| *JM-1L-DI-wide* | -0.008 (0.018) | -0.003 (0.003) | 0.047 | 0.244 | 0.230 |
| *FCS-1L-DI-wide* | -0.012 (0.020) | -0.002 (0.003) | 0.047 | 0.245 | 0.230 |
| *JM-2L-wide* | -0.013 (0.020) | -0.002 (0.004) | 0.042 | 0.245 | 0.229 |
| *FCS-2L-wide* | -0.012 (0.018) | -0.002 (0.003) | 0.042 | 0.248 | 0.227 |
| *SMC-JM-2L-DI* | -0.012 (0.019) | -0.002 (0.004) | 0.028 | 0.252 | 0.229 |
| *SMC-SM-2L-DI* | -0.009 (0.020) | -0.002 (0.004) | 0.029 | 0.295 | 0.229 |
| *SMC-JM-3L* | -0.013 (0.018) | -0.002 (0.003) | 0.040 | 0.240 | 0.227 |

Table 4: Point estimate (and standard error) for the main effect and the interaction effect, and point estimates for the variance components at levels 3, 2 and 1, from 11 MI approaches applied to the CATS data analysis in analysis model 2.

| Method | Main effect (SE) | Interaction effect (SE) | Level 3 variance component | Level 2 variance component | Level 1 variance component |
|---|---|---|---|---|---|
| *JM-1L-DI-wide* | -0.020 (0.006) | 0.007 (0.007) | 0.047 | 0.244 | 0.230 |
| *JM-1L-DI-wide-JAV* | -0.020 (0.007) | 0.019 (0.008) | 0.046 | 0.250 | 0.230 |
| *FCS-1L-DI-wide* | -0.022 (0.007) | 0.008 (0.006) | 0.047 | 0.245 | 0.230 |
| *FCS-1L-DI-wide-passive_c* | -0.022 (0.006) | 0.010 (0.007) | 0.046 | 0.239 | 0.229 |
| *FCS-1L-DI-wide-passive_all* | -0.021 (0.007) | 0.010 (0.006) | 0.046 | 0.239 | 0.227 |
| *JM-2L-wide-JAV* | -0.020 (0.007) | 0.017 (0.008) | 0.042 | 0.244 | 0.229 |
| *FCS-2L-wide-passive_c* | -0.021 (0.007) | 0.007 (0.006) | 0.044 | 0.247 | 0.229 |
| *FCS-2L-wide-passive_all* | -0.021 (0.007) | 0.008 (0.007) | 0.043 | 0.246 | 0.229 |
| *SMC-JM-2L-DI* | -0.024 (0.007) | 0.006 (0.006) | 0.028 | 0.253 | 0.229 |
| *SMC-SM-2L-DI* | -0.020 (0.008) | 0.007 (0.006) | 0.029 | 0.300 | 0.228 |
| *SMC-JM-3L* | -0.022 (0.006) | 0.007 (0.007) | 0.040 | 0.239 | 0.226 |

Table 5: Point estimate (and standard error) for the main effect and the quadratic effect, and point estimates for the variance components at levels 3, 2 and 1, from 9 MI approaches applied to the CATS data analysis in analysis model 3.

| Method | Main effect (SE) | Quadratic effect (SE) | Level 3 variance component | Level 2 variance component | Level 1 variance component |
|---|---|---|---|---|---|
| *JM-1L-DI-wide* | -0.017 (0.010) | -0.0014 (0.002) | 0.046 | 0.243 | 0.231 |
| *JM-1L-DI-wide-JAV* | -0.012 (0.013) | -0.0029 (0.003) | 0.045 | 0.249 | 0.231 |
| *FCS-1L-DI-wide* | -0.018 (0.012) | -0.0016 (0.003) | 0.047 | 0.245 | 0.230 |
| *FCS-1L-DI-wide-passive* | -0.015 (0.013) | -0.0022 (0.003) | 0.050 | 0.246 | 0.232 |
| *JM-2L-wide-JAV* | -0.012 (0.012) | -0.0028 (0.003) | 0.041 | 0.243 | 0.229 |
| *FCS-2L-wide-passive* | -0.013 (0.011) | -0.0025 (0.003) | 0.042 | 0.245 | 0.227 |
| *SMC-JM-2L-DI* | -0.013 (0.011) | -0.0025(0.003) | 0.027 | 0.251 | 0.228 |
| *SMC-SM-2L-DI* | -0.012 (0.012) | -0.0030 (0.003) | 0.029 | 0.295 | 0.228 |
| *SMC-JM-3L* | -0.013 (0.012) | -0.0026 (0.003) | 0.040 | 0.240 | 0.227 |

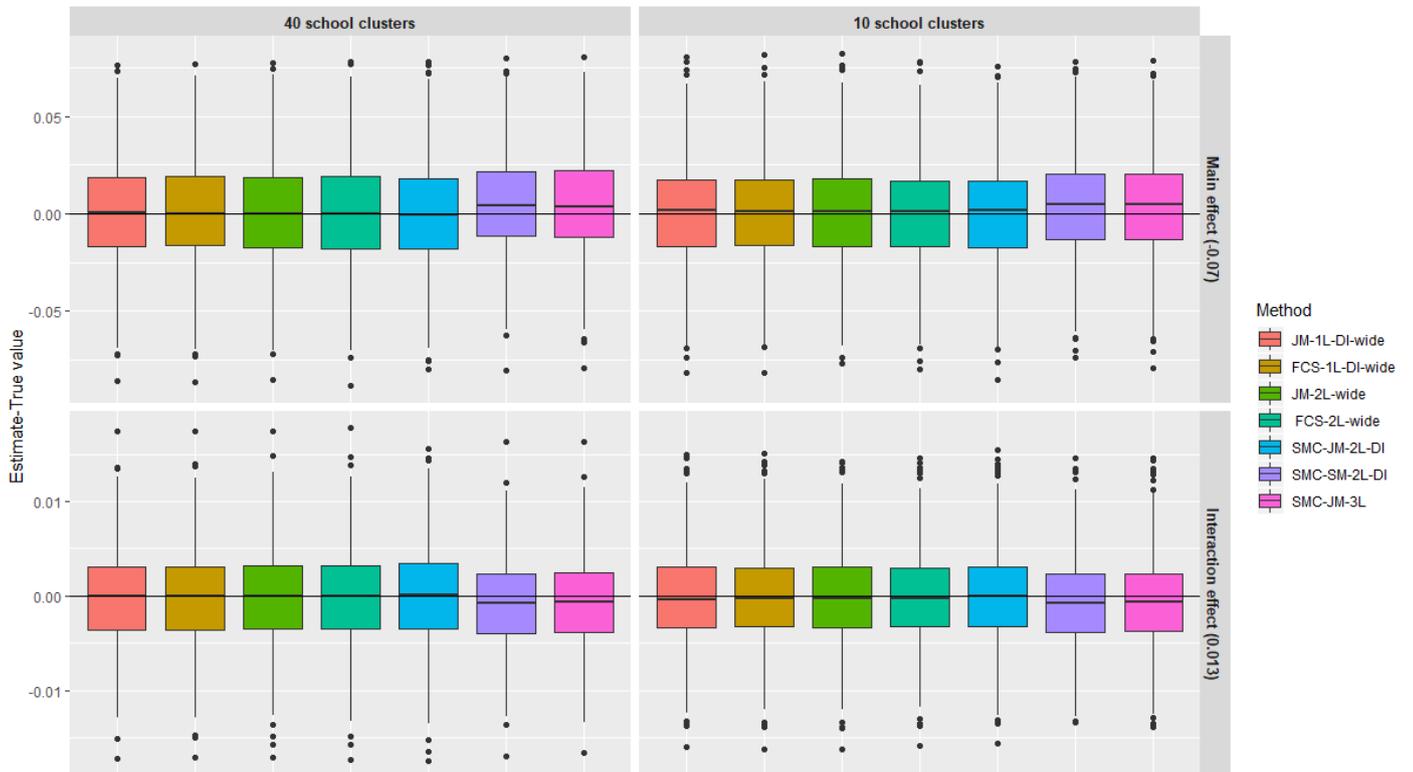

Figure 1: Distribution of the bias in the estimated regression coefficient for the main effect ($\beta_1$, $true\ value = -0.07$) and the interaction effect ($\beta_3$, $true\ value = 0.013$) in analysis model 1, across the 1000 simulated datasets for the 7 multiple imputation (MI) approaches under two scenarios for number of higher level clusters ( 40 school clusters and 10 school clusters) when data are missing at random with relationships based on the CATS data (MAR-CATS)

The lower and upper margins of the boxes represent the $25^{th}$ ($Q_1$) and the $75^{th}$ ($Q_3$) percentiles of the distribution respectively. The whiskers extend to $Q_1-1.5*(Q_3-Q_1)$ at the bottom and $Q_3+1.5*(Q_3-Q_1)$ at the top.

The following abbreviations are used to denote different MI methods, e.g., DI: dummy indicators, FCS: fully conditional specification, JM: joint modelling, SM: sequential modelling, SMC: substantive model compatible.

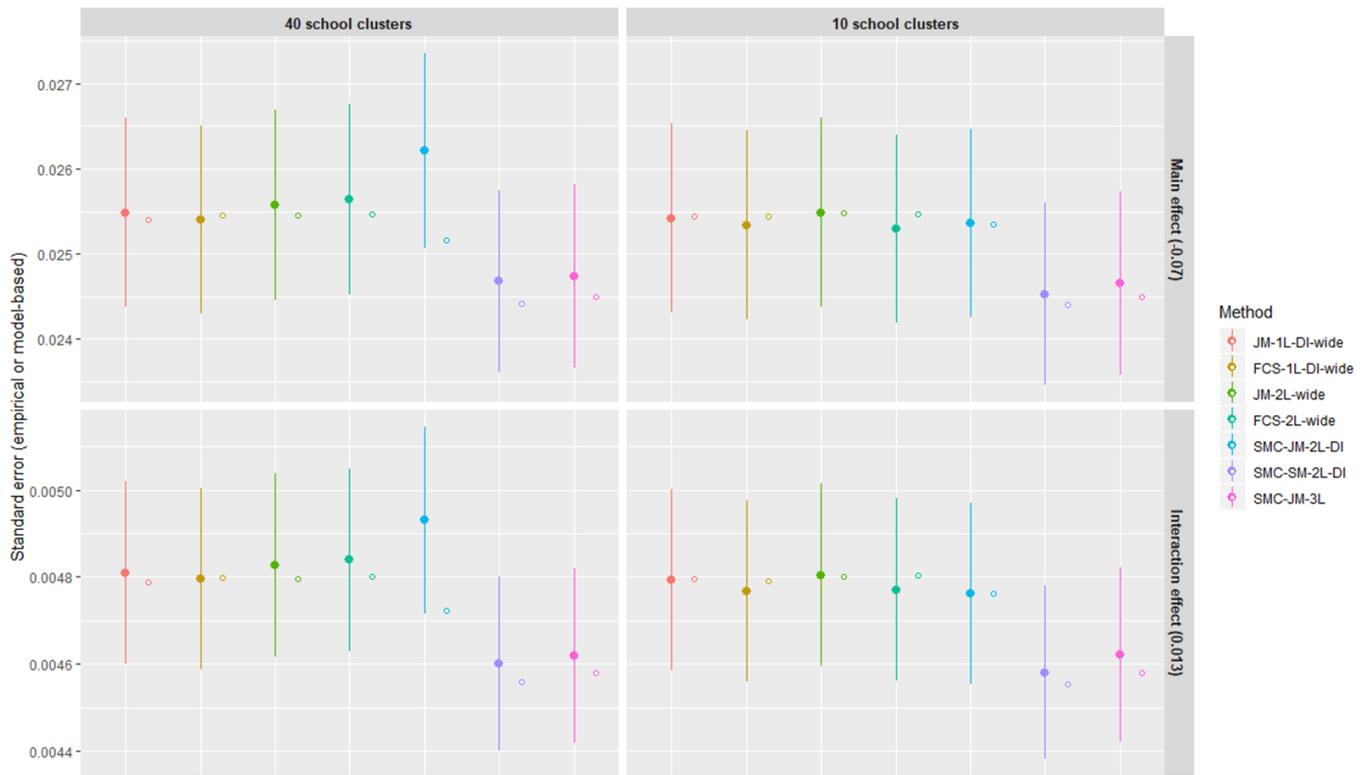

Figure 2: Empirical standard error (filled circles with error bars showing ±1.96× Monte Carlo standard errors) and average model-based standard error (hollow circles) in analysis model 1 from 1000 simulated datasets, for the 7 multiple imputation (MI) approaches under two scenarios for number of higher level clusters (40 school clusters and 10 school clusters) when data are missing at random with relationships based on the CATS data (MAR-CATS)

The following abbreviations are used to denote different MI methods, e.g., DI: dummy indicators, FCS: fully conditional specification, JM: joint modelling, SM: sequential modelling, SMC: substantive model compatible

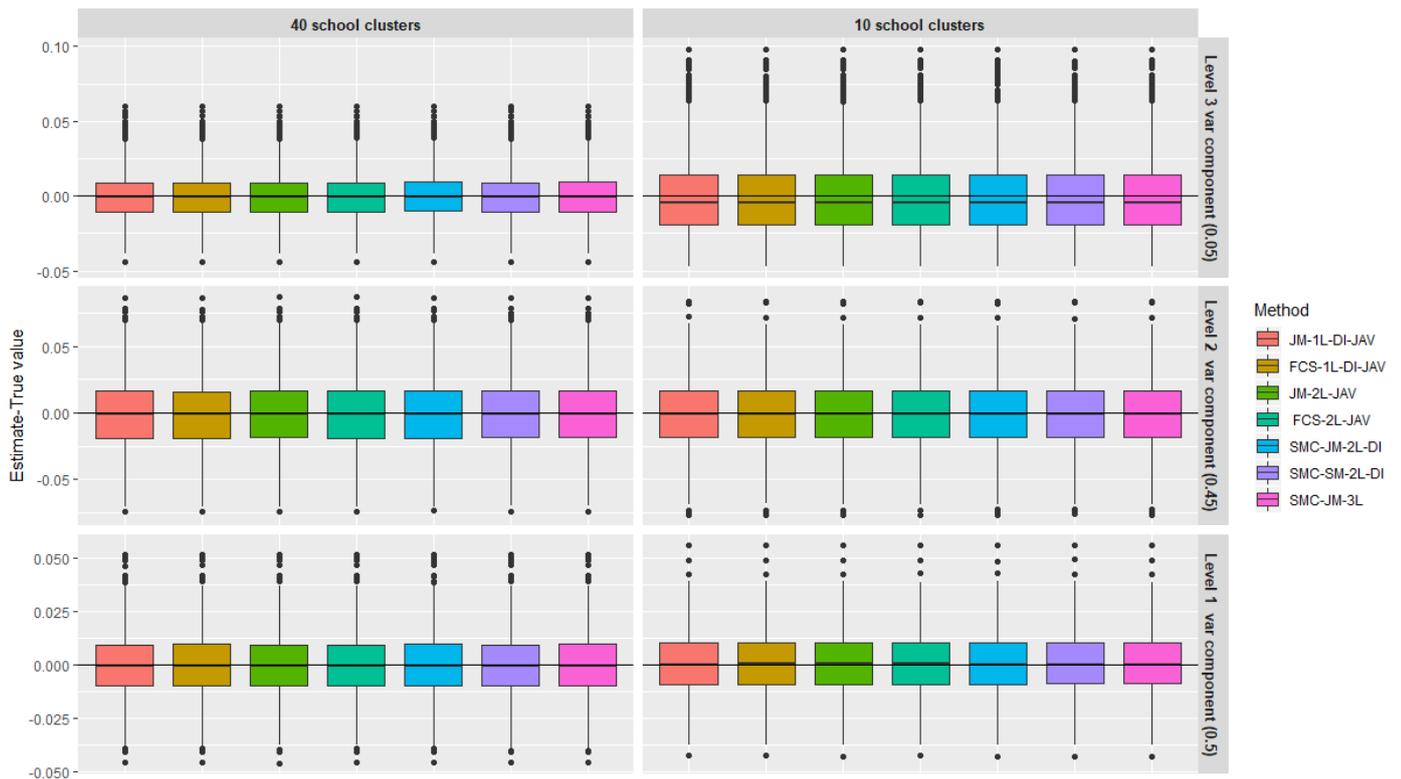

Figure 3: Distribution of the bias in the estimated variance components at level 1, 2 and 3 in analysis model 1 across the 1000 simulated datasets for the 7 multiple imputation (MI) approaches under two scenarios for number of higher level clusters (40 school clusters and 10 school clusters) when data are missing at random with relationships based on the CATS data (MAR-CATS)

The following abbreviations are used to denote different MI methods, e.g., DI: dummy indicators, FCS: fully conditional specification, JM: joint modelling, SM: sequential modelling, SMC: substantive model compatible.

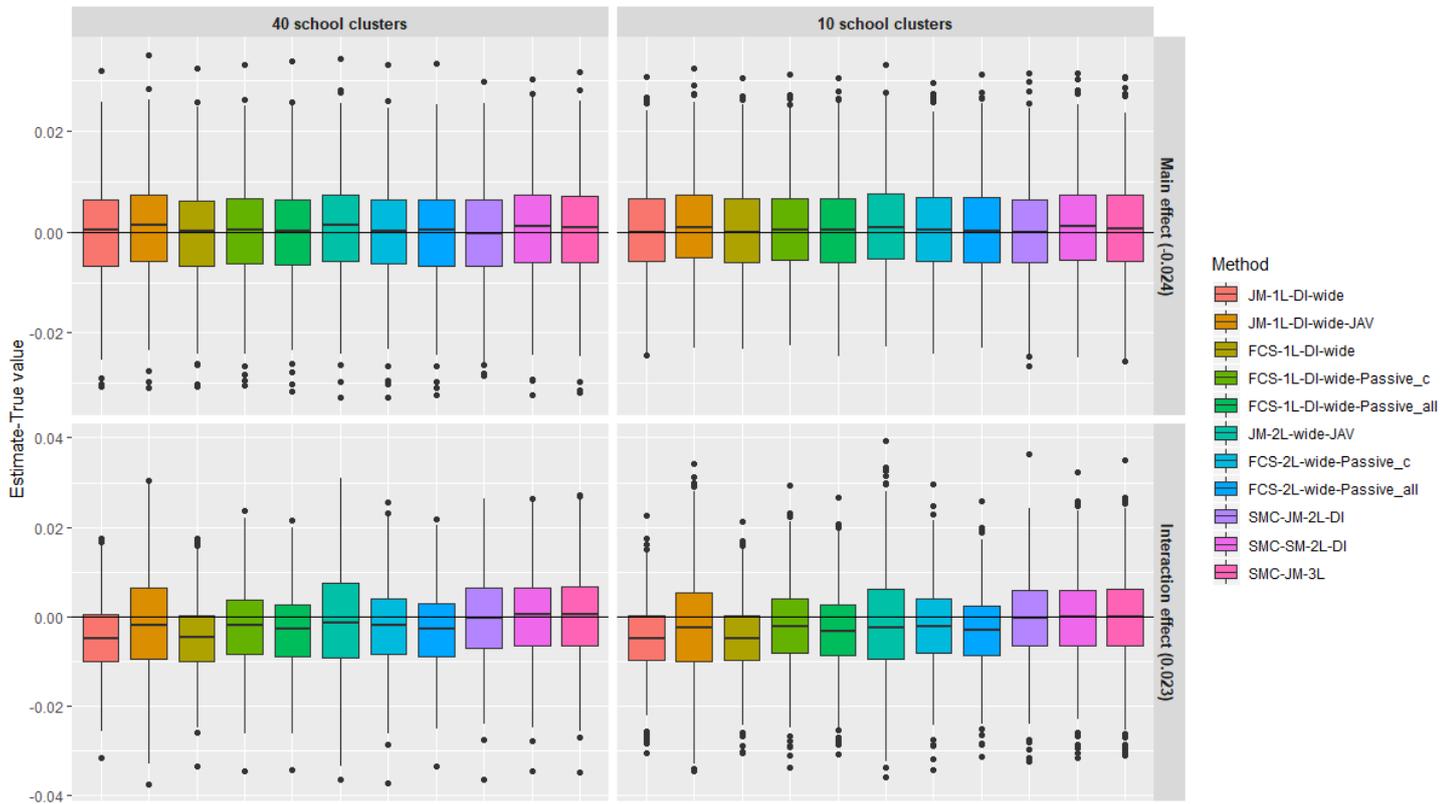

Figure 4: Distribution of the bias in the estimated regression coefficient for the main effect ($\beta_1$, true value=-0.024) and the interaction effect ($\beta_3$, true value=0.023) in analysis model 2, across the 1000 simulated datasets for the 11 multiple imputation (MI) approaches under two scenarios for number of higher level clusters (40 school clusters and 10 school clusters) when data are missing at random (MAR-CATS)

The lower and upper margins of the boxes represent the 25th (Q1) and the 75th (Q3) percentiles of the distribution respectively. The whiskers extend to Q1-1.5*(Q3- Q1) at the bottom and Q3 +1.5*(Q3-Q1) at the top.

The following abbreviations are used to denote different MI methods, e.g., DI: dummy indicators, FCS: fully conditional specification, JM: joint modelling, JAV: Just another variable, Passive_all: passive imputation where all of the possible interactions between depressive symptoms, NAPLAN scores and SES are included as predictors in the imputation model, Passive_c: passive imputation where just the interaction between depressive symptoms and the NAPLAN scores at the current wave and SES are included as predictors in the imputation model), SM: sequential modelling, SMC: substantive model compatible.

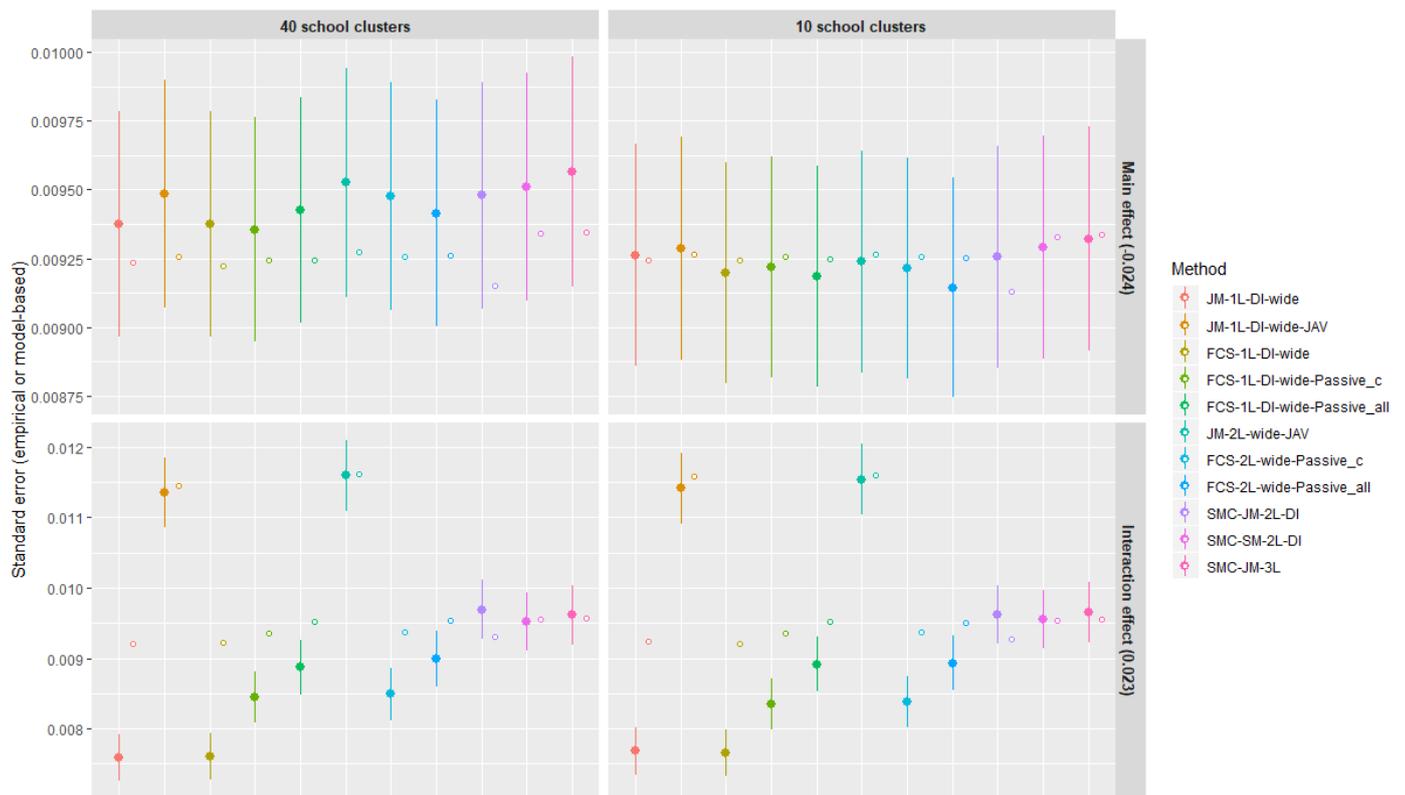

Figure 5: Empirical standard errors (filled circles with error bars showing ±1.96× Monte Carlo standard errors) and average model-based standard errors (hollow circles) in analysis model 2 from 1000 simulated datasets, for the 11 multiple imputation (MI) approaches under two scenarios for number of higher level clusters (40 school clusters and 10 school clusters) when data are missing at random (MAR-CATS)

The following abbreviations are used to denote different MI methods, e.g., DI: dummy indicators, FCS: fully conditional specification, JM: joint modelling, JAV: Just another variable, Passive_all: passive imputation (with all the interactions as predictors in imputing incomplete depressive symptoms), Passive_c: passive imputation (with just the interaction between the NAPLAN scores at the current wave and SES as predictors in imputing incomplete previous wave depressive symptoms), SM: sequential modelling, SMC: substantive model compatible.

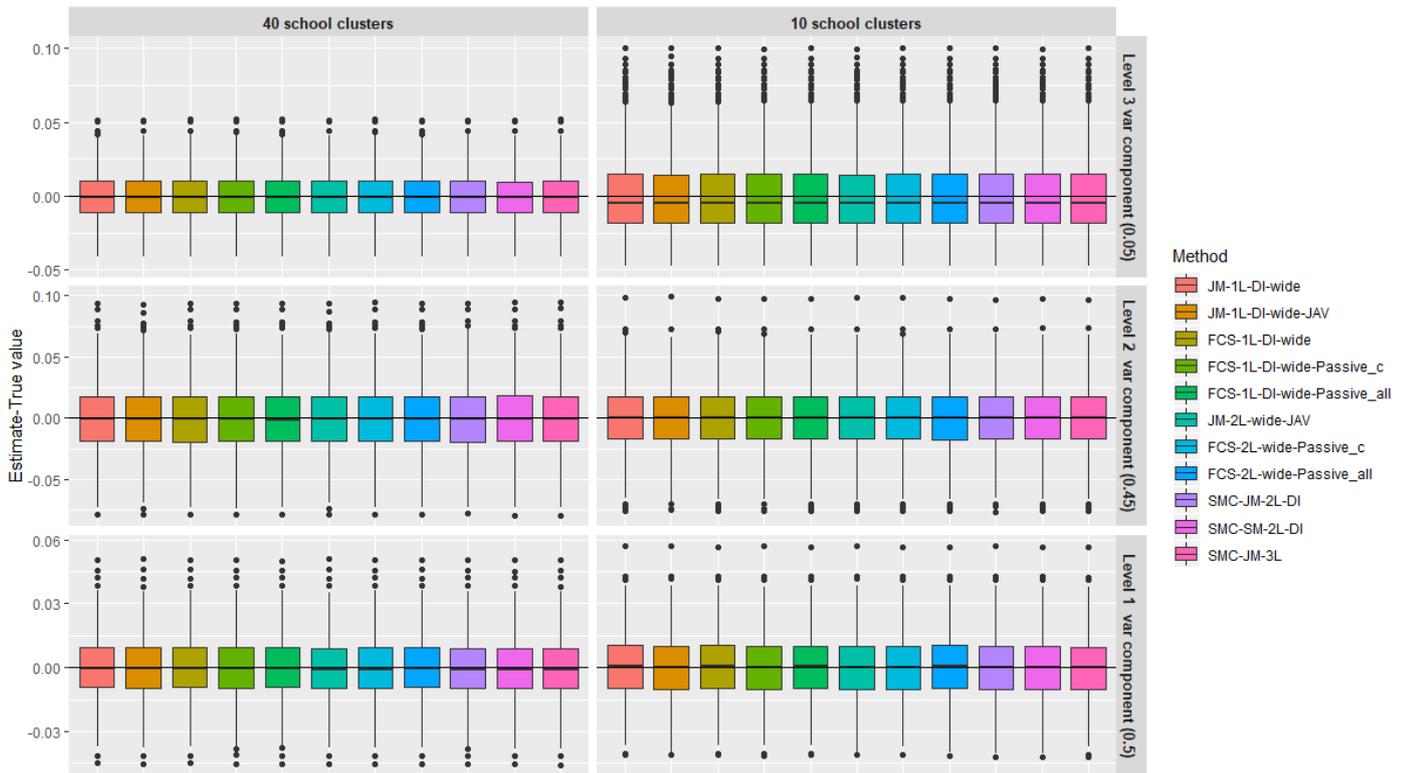

Figure 6: Distribution of the bias in the estimated variance components at level 1, 2 and 3 in analysis model 2 across the 1000 simulated datasets for the 11 multiple imputation (MI) approaches under two scenarios for number of higher level clusters (40 school clusters and 10 school clusters) when data are missing at random (MAR-CATS)

The following abbreviations are used to denote different MI methods, e.g., DI: dummy indicators, FCS: fully conditional specification, JM: joint modelling, JAV: Just another variable, Passive_all: passive imputation (with all the interactions as predictors in imputing incomplete depressive symptoms), Passive_c: passive imputation (with just the interaction between the NAPLAN scores at the current wave and SES as predictors in imputing incomplete previous wave depressive symptoms), SM: sequential modelling, SMC: substantive model compatible.

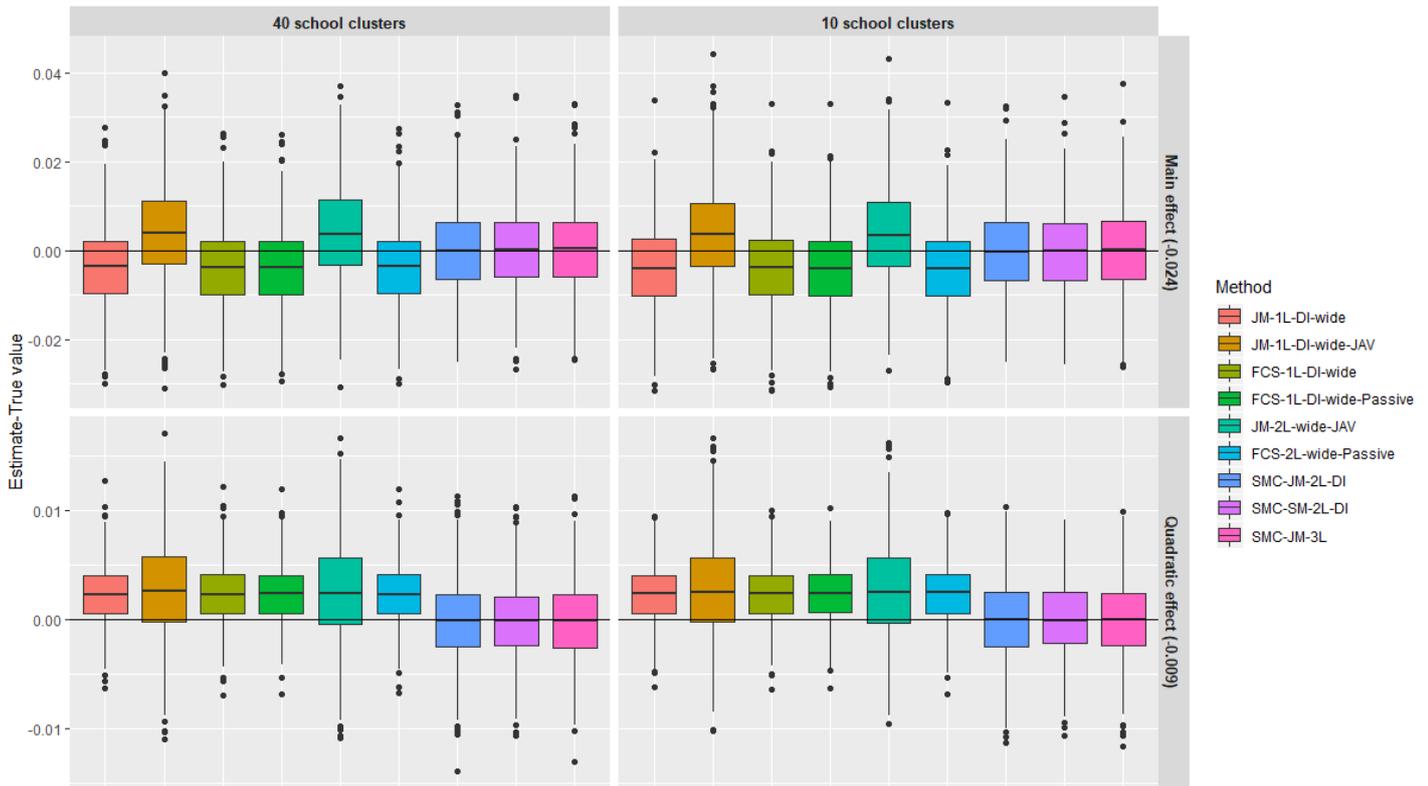

Figure 7: Distribution of the bias in the estimated regression coefficient for the main effect ($\beta_1$, true value=-0.024) and the quadratic effect ($\beta_3$, true value=-0.009) in analysis model 3, across the 1000 simulated datasets for the 9 multiple imputation (MI) approaches under two scenarios for number of higher level clusters ( 40 school clusters and 10 school clusters) when data are missing at random (MAR-CATS). The lower and upper margins of the boxes represent the 25th (Q1) and the 75th (Q3) percentiles of the distribution respectively. The whiskers extend to Q1-1.5*(Q3- Q1) at the bottom and Q3 +1.5*(Q3- Q1) at the top.

The following abbreviations are used to denote different MI methods, e.g., DI: dummy indicators, FCS: fully conditional specification, JM: joint modelling, SM: sequential modelling, SMC: substantive model compatible.

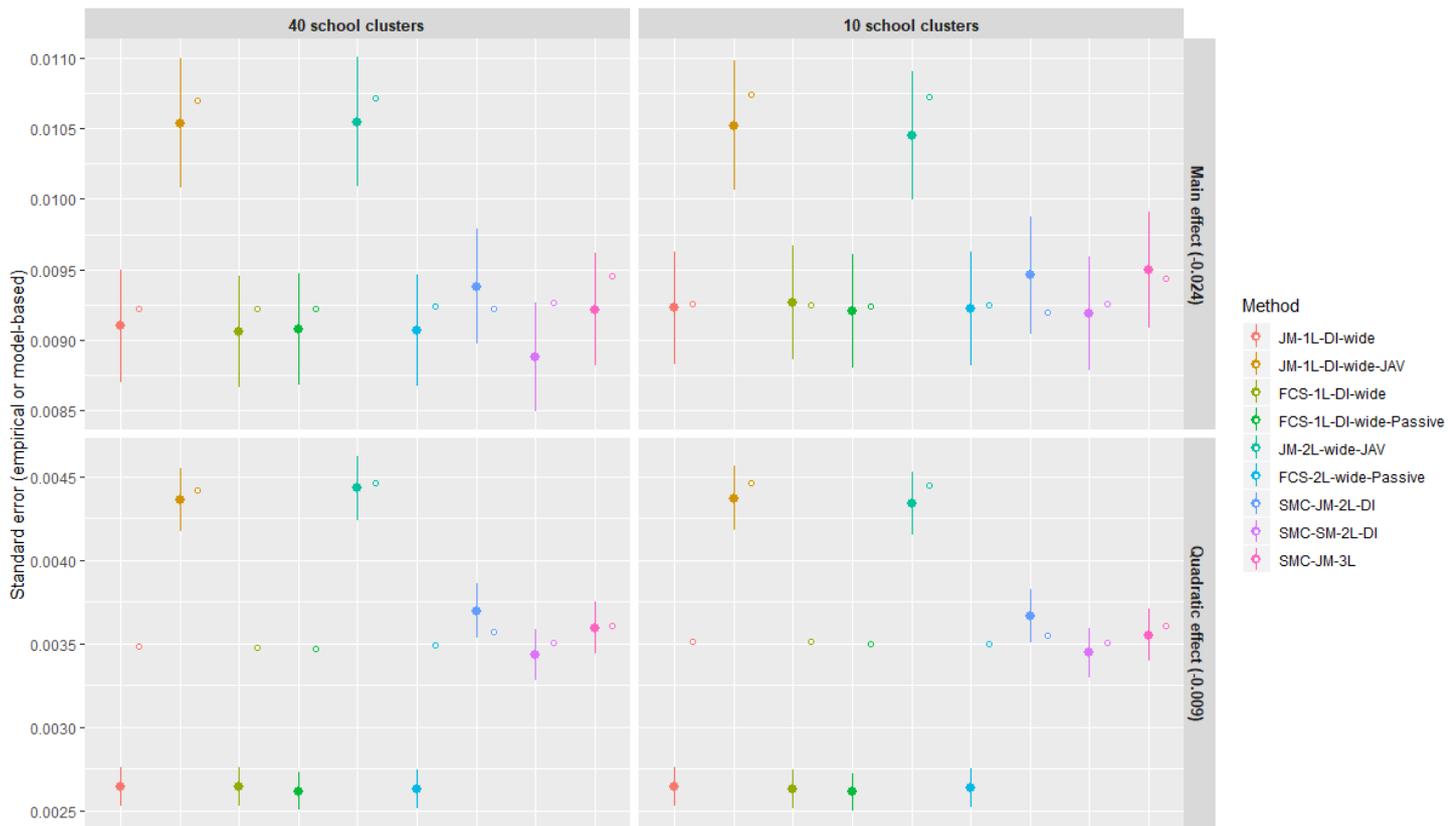

Figure 8: Empirical standard errors (filled circles with error bars showing ±1.96× Monte Carlo standard errors) and average model-based standard errors (hollow circles) in analysis model 3 from 1000 simulated datasets, for the 11 multiple imputation (MI) approaches under two scenarios for number of higher level clusters (40 school clusters and 10 school clusters) when data are missing at random (MAR-CATS)

The following abbreviations are used to denote different MI methods, e.g., DI: dummy indicators, FCS: fully conditional specification, JM: joint modelling, SM: sequential modelling, SMC: substantive model compatible.

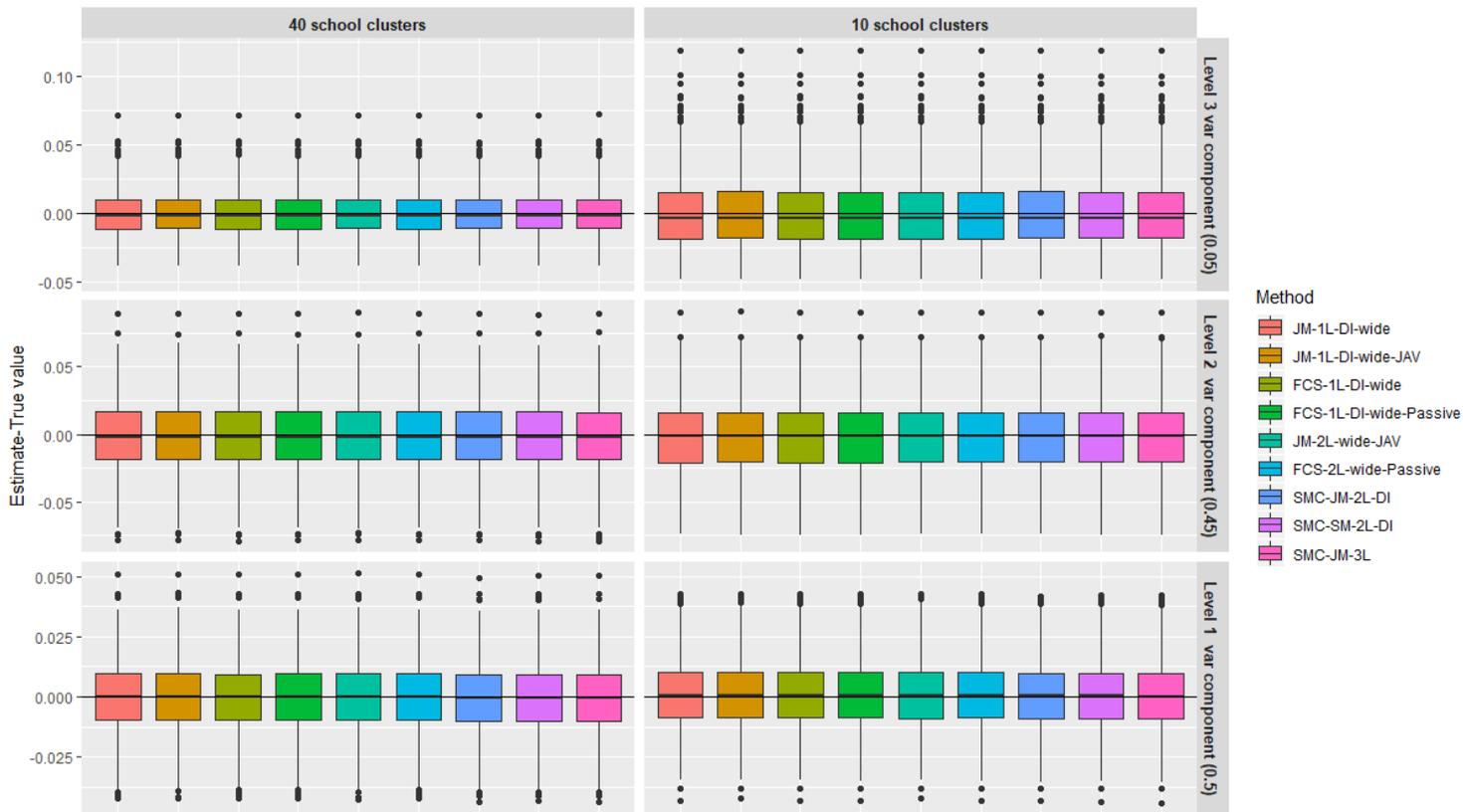

Figure 9: Distribution of the bias in the estimated variance components at level 1, 2 and 3 in analysis model 3 across the 1000 simulated datasets for the 9 multiple imputation (MI) approaches under two scenarios for number of higher level clusters (40 school clusters and 10 school clusters) when data are missing at random (MAR-inflated)

The following abbreviations are used to denote different MI methods, e.g., DI: dummy indicators, FCS: fully conditional specification, JM: joint modelling, SM: sequential modelling, SMC: substantive model compatible.

# Additional Files

Table of Contents

# Supplement A: Parameter values used in the simulation study for complete and missing data generation

Table S1: Parameter values used in the simulation study

| Variable generated | Parameter | | Value |
|---|---|---|---|
| Child's age at wave 1 | Lower limit | $a$ | 7 |
| | Upper limit | $b$ | 10 |
| Child's sex | Proportion of females | $\lambda\%$ | 0.5 |
| Child's SES value at wave 1 | | Mean | 0 |
| | | Standard deviation | 1 |
| Standardized NAPLAN numeracy scores at wave 1 | Constant | $\eta_0$ | -0.74 |
| | Sex | $\eta_1$ | 0.23 |
| | Age | $\eta_2$ | 0.07 |
| | SES at wave 1 | $\eta_3$ | 0.22 |
| | Error term | Standard deviation $\sigma_\psi$ | 1 |
| Child's depressive symptom scores at waves 2-7 | Constant | $\delta_0$ | -0.7 |
| | Age | $\delta_1$ | 0.1 |
| | Sex | $\delta_2$ | -0.46 |
| | NAPLAN numeracy scores at wave 1 | $\delta_3$ | -0.01 |
| | SES at wave 1 | $\delta_4$ | -0.22 |
| | Wave | $\delta_5$ | 0.02 |
| | Random effects- Level 3 | Standard deviation $\sigma_{u_3}$ | 0.1 |
| | Level 2 | Standard deviation $\sigma_{u_2}$ | 0.9 |
| | Error term | Standard deviation $\sigma_\varphi$ | 1.5 |
| Child's standardized NAPLAN numeracy scores at waves 3,5 and 7 | Constant | $\beta_0$ | 2.0 |
| | Depressive symptom score at previous wave | $\beta_1$ | See main text |
| | Wave | $\beta_2$ | -0.01 |
| | Interaction term/ the quadratic effect | $\beta_3$ | See main text |
| | NAPLAN numeracy scores at wave 1 | $\beta_4$ | 0.71 |

| | | | | | |
|---|---|---|---|---|---|
| | Sex | | $\beta_5$ | | 0.14 |
| | SES at wave 1 7 | | $\beta_{6,0}$ | | -0.01 |
| | Age | | $\beta_{6,1}$ | | -0.20 |
| | Random effects-Level 3 | | Standard deviation $\sigma_3$ | | 0.2 |
| | Level 2 | | Standard deviation $\sigma_2$ | | 0.7 |
| | Error term | | Standard deviation $\sigma_1$ | | 0.7 |
| Overall child behaviour reported by SDQ at waves 2,4 and 6 | Constant | | $\gamma_0$ | | 16.2 |
| | Depressive symptom score at concurrent wave | | $\gamma_1$ | | 2.5 |
| | Wave | | $\gamma_2$ | | -0.1 |
| | Random effects-Level 3 | | Standard deviation $\sigma_{v_3}$ | | 0.6 |
| | Level 2 | | Standard deviation $\sigma_{v_2}$ | | 4.1 |
| | Error term | | Standard deviation $\sigma_\epsilon$ | | 2.8 |

SES: Socio-Economic Status, NAPLAN: National Assessment Program - Literacy and Numeracy, SDQ: Strengths and Difficulties Questionnaire.

Note: For some of the simulation scenarios 1-3 datasets were omitted and replaced with different simulated datasets due to non-convergence in SMC-JM-3L. The replications omitted are noted in the simulation code provided.

## Supplement B: Performance of multiple imputation (MI) methods under analysis model 1

Table S2: Performance of the 7 multiple imputation (MI) methods for estimating the main effect ($\beta_1 = -0.07$) and the interaction effect ($\beta_3 = 0.013$) in the analysis model 1 when data are missing at random with dependencies based on the CATS data (MAR-CATS)

| No of higher-level clusters | Method | Main effect (-0.07) | | | | | | Interaction effect (0.013) | | | | | |
|---|---|---|---|---|---|---|---|---|---|---|---|---|---|
| | | Average Estimate | Bias | Relative Bias (%) | Emp SE | Model SE | Coverage | Average Estimate | Bias | Relative Bias (%) | Emp SE | Model SE | Coverage |
| 40 clusters | JM-1L-DI-wide | -0.069 | 0.0008 | -1.2 | 0.0256 | 0.0254 | 94.6% | 0.013 | -0.0001 | -1.0 | 0.0048 | 0.0048 | 95.2% |
| | FCS-1L-DI-wide | -0.069 | 0.0010 | -1.4 | 0.0255 | 0.0255 | 95.3% | 0.013 | -0.0002 | -1.2 | 0.0048 | 0.0048 | 95.6% |
| | JM-2L-wide | -0.069 | 0.0006 | -0.9 | 0.0256 | 0.0255 | 95.0% | 0.013 | -0.0001 | -0.4 | 0.0048 | 0.0048 | 94.5% |
| | FCS-2L-wide | -0.069 | 0.0005 | -0.7 | 0.0256 | 0.0255 | 94.9% | 0.013 | -0.0001 | -0.4 | 0.0048 | 0.0048 | 95.0% |
| | SMC-JM-2L-DI | -0.070 | 0.0002 | -0.2 | 0.0261 | 0.0252 | 94.1% | 0.013 | 0.0000 | -0.4 | 0.0049 | 0.0047 | 94.4% |
| | SMC-SM-2L-DI | -0.065 | 0.0047 | -6.7 | 0.0247 | 0.0244 | 94.1% | 0.012 | -0.0008 | -5.8 | 0.0046 | 0.0046 | 94.7% |
| | SMC-JM-3L | -0.066 | 0.0041 | -5.9 | 0.0247 | 0.0245 | 94.2% | 0.012 | -0.0007 | -5.0 | 0.0046 | 0.0046 | 94.2% |

| No of higher-level clusters | Method | | | | | | | | | | | | |
|---|---|---|---|---|---|---|---|---|---|---|---|---|---|
| 10 clusters | JM-1L-DI-wide | -0.069 | 0.0008 | -1.1 | 0.0254 | 0.0254 | 94.7% | 0.013 | -0.0001 | -1.0 | 0.0048 | 0.0048 | 94.6% |
| | FCS-1L-DI-wide | -0.069 | 0.0008 | -1.2 | 0.0253 | 0.0254 | 94.5% | 0.013 | -0.0001 | -1.1 | 0.0048 | 0.0048 | 94.5% |
| | JM-2L-wide | -0.069 | 0.0008 | -1.2 | 0.0255 | 0.0255 | 94.6% | 0.013 | -0.0001 | -1.1 | 0.0048 | 0.0048 | 94.5% |
| | FCS-2L-wide | -0.069 | 0.0007 | -1.0 | 0.0253 | 0.0255 | 94.7% | 0.013 | -0.0001 | -0.9 | 0.0048 | 0.0048 | 94.8% |
| | SMC-JM-2L-DI | -0.070 | 0.0001 | -0.1 | 0.0254 | 0.0254 | 94.9% | 0.013 | 0.0000 | -0.2 | 0.0048 | 0.0048 | 94.5% |
| | SMC-SM-2L-DI | -0.065 | 0.0045 | -6.5 | 0.0245 | 0.0244 | 94.1% | 0.012 | -0.0008 | -5.8 | 0.0046 | 0.0046 | 94.0% |
| | SMC-JM-3L | -0.066 | 0.0041 | -5.8 | 0.0247 | 0.0245 | 94.0% | 0.012 | -0.0007 | -5.2 | 0.0046 | 0.0046 | 94.2% |

Table S3: Performance of the 7 multiple imputation (MI) methods in estimating the variance components (VC) at level 3, 2 and 1 in the analysis model 1 when data are missing at random with dependencies based on the CATS data (MAR-CATS)

| No of higher-level clusters | Method | Level 3 VC (0.05) | | | Level 2 VC (0.45) | | | Level 1 VC (0.5) | | |
|---|---|---|---|---|---|---|---|---|---|---|
| | | Bias | Relative Bias (%) | Emp SE | Bias | Relative Bias (%) | Emp SE | Bias | Relative Bias (%) | Emp SE |
| 40 clusters | JM-1L-DI-wide | 0.0000 | 0.09 | 0.016 | -0.0004 | -0.09 | 0.026 | 0.0002 | 0.03 | 0.015 |
| | FCS-1L-DI-wide | 0.0000 | 0.06 | 0.016 | -0.0004 | -0.09 | 0.026 | 0.0002 | 0.03 | 0.015 |
| | JM-2L-wide | 0.0000 | 0.04 | 0.016 | -0.0004 | -0.08 | 0.026 | 0.0001 | 0.03 | 0.015 |
| | FCS-2L-wide | 0.0000 | 0.02 | 0.016 | -0.0004 | -0.08 | 0.026 | 0.0001 | 0.03 | 0.015 |
| | SMC-JM-2L-DI | 0.0000 | 0.04 | 0.016 | -0.0004 | -0.09 | 0.026 | 0.0001 | 0.03 | 0.015 |
| | SMC-SM-2L-DI | 0.0000 | -0.02 | 0.016 | -0.0003 | -0.07 | 0.026 | 0.0002 | 0.04 | 0.015 |
| | SMC-JM-3L | 0.0000 | 0.03 | 0.016 | -0.0003 | -0.08 | 0.026 | 0.0002 | 0.04 | 0.015 |
| 10 clusters | JM-1L-DI-wide | -0.0006 | -1.13 | 0.026 | -0.0008 | -0.17 | 0.026 | 0.0004 | 0.07 | 0.014 |
| | FCS-1L-DI-wide | -0.0006 | -1.11 | 0.026 | -0.0007 | -0.16 | 0.026 | 0.0004 | 0.07 | 0.014 |
| | JM-2L-wide | -0.0006 | -1.13 | 0.026 | -0.0007 | -0.16 | 0.026 | 0.0004 | 0.07 | 0.014 |
| | FCS-2L-wide | -0.0006 | -1.11 | 0.026 | -0.0007 | -0.17 | 0.026 | 0.0004 | 0.07 | 0.014 |
| | SMC-JM-2L-DI | -0.0005 | -1.08 | 0.026 | -0.0008 | -0.17 | 0.026 | 0.0004 | 0.07 | 0.014 |
| | SMC-SM-2L-DI | -0.0006 | -1.14 | 0.026 | -0.0007 | -0.16 | 0.026 | 0.0004 | 0.08 | 0.014 |
| | SMC-JM-3L | -0.0006 | -1.11 | 0.026 | -0.0007 | -0.16 | 0.026 | 0.0004 | 0.08 | 0.014 |

Table S4: Performance of the 7 multiple imputation (MI) methods for estimating the main effect ($\beta_1 = -0.07$) and the interaction effect ($\beta_3 = 0.013$) in the analysis model 1 when data are missing at random with inflated dependencies (MAR-inflated)

| | Method | Main effect (-0.07) | Interaction effect (0.013) |
|---|---|---|---|

| No of higher-level clusters | | Average Estimate | Bias | Relative Bias (%) | Emp SE | Model SE | Coverage | Average Estimate | Bias | Relative Bias (%) | Emp SE | Model SE | Coverage |
|---|---|---|---|---|---|---|---|---|---|---|---|---|---|
| 40 clusters | JM-1L-DI-wide | -0.070 | 0.0001 | -0.2 | 0.0244 | 0.0255 | 95.5% | 0.013 | 0.0000 | -0.2 | 0.0046 | 0.0048 | 95.7% |
| | FCS-1L-DI-wide | -0.070 | 0.0001 | -0.1 | 0.0244 | 0.0255 | 95.7% | 0.013 | 0.0000 | -0.3 | 0.0046 | 0.0048 | 95.7% |
| | JM-2L-wide | -0.070 | -0.0005 | 0.7 | 0.0245 | 0.0256 | 95.5% | 0.013 | 0.0001 | 0.7 | 0.0046 | 0.0048 | 95.9% |
| | FCS-2L-wide | -0.070 | -0.0005 | 0.7 | 0.0244 | 0.0256 | 95.8% | 0.013 | 0.0001 | 0.7 | 0.0046 | 0.0048 | 95.6% |
| | SMC-JM-2L-DI | -0.072 | -0.0016 | 2.3 | 0.0252 | 0.0253 | 95.3% | 0.013 | 0.0002 | 1.7 | 0.0047 | 0.0047 | 94.2% |
| | SMC-SM-2L-DI | -0.065 | 0.0047 | -6.7 | 0.0235 | 0.0244 | 94.7% | 0.012 | -0.0007 | -5.6 | 0.0044 | 0.0045 | 95.1% |
| | SMC-JM-3L | -0.066 | 0.0040 | -5.7 | 0.0237 | 0.0245 | 95.1% | 0.012 | -0.0006 | -4.6 | 0.0044 | 0.0046 | 95.3% |
| 10 clusters | JM-1L-DI-wide | -0.071 | -0.0011 | 1.6 | 0.0254 | 0.0256 | 94.8% | 0.013 | 0.0003 | 2.1 | 0.0047 | 0.0048 | 95.4% |
| | FCS-1L-DI-wide | -0.071 | -0.0010 | 1.4 | 0.0254 | 0.0256 | 95.2% | 0.013 | 0.0002 | 1.8 | 0.0047 | 0.0048 | 95.4% |
| | JM-2L-wide | -0.071 | -0.0010 | 1.4 | 0.0255 | 0.0256 | 95.2% | 0.013 | 0.0002 | 1.8 | 0.0047 | 0.0048 | 95.6% |
| | FCS-2L-wide | -0.071 | -0.0011 | 1.5 | 0.0253 | 0.0256 | 95.4% | 0.013 | 0.0003 | 2.0 | 0.0047 | 0.0048 | 95.7% |
| | SMC-JM-2L-DI | -0.072 | -0.0023 | 3.3 | 0.0252 | 0.0255 | 95.0% | 0.013 | 0.0004 | 3.1 | 0.0047 | 0.0048 | 96.1% |
| | SMC-SM-2L-DI | -0.066 | 0.0040 | -5.7 | 0.0242 | 0.0243 | 94.8% | 0.012 | -0.0005 | -4.0 | 0.0045 | 0.0045 | 95.4% |
| | SMC-JM-3L | -0.067 | 0.0034 | -4.8 | 0.0243 | 0.0245 | 95.1% | 0.013 | -0.0004 | -3.3 | 0.0045 | 0.0046 | 95.7% |

Table S5: Performance of the 7 multiple imputation (MI) methods in estimating the variance components (VC) at level 3, 2 and 1 in the analysis model 1 when data are missing at random with inflated dependencies (MAR-inflated)

| No of higher-level clusters | Method | Level 3 VC (0.05) | | | Level 2 VC (0.45) | | | Level 1 VC (0.5) | | |
|---|---|---|---|---|---|---|---|---|---|---|
| | | Bias | Relative Bias (%) | Emp SE | Bias | Relative Bias (%) | Emp SE | Bias | Relative Bias (%) | Emp SE |
| 40 clusters | JM-1L-DI-wide | 0.0006 | 1.26 | 0.016 | -0.0026 | -0.57 | 0.026 | 0.0001 | 0.02 | 0.015 |
| | FCS-1L-DI-wide | 0.0006 | 1.26 | 0.016 | -0.0026 | -0.57 | 0.026 | 0.0001 | 0.02 | 0.015 |
| | JM-2L-wide | 0.0006 | 1.25 | 0.016 | -0.0026 | -0.57 | 0.026 | 0.0001 | 0.02 | 0.015 |
| | FCS-2L-wide | 0.0006 | 1.24 | 0.016 | -0.0026 | -0.57 | 0.026 | 0.0001 | 0.02 | 0.015 |
| | SMC-JM-2L-DI | 0.0006 | 1.28 | 0.016 | -0.0026 | -0.58 | 0.026 | 0.0001 | 0.02 | 0.015 |
| | SMC-SM-2L-DI | 0.0006 | 1.18 | 0.016 | -0.0025 | -0.56 | 0.026 | 0.0002 | 0.03 | 0.015 |
| | SMC-JM-3L | 0.0006 | 1.24 | 0.016 | -0.0026 | -0.57 | 0.026 | 0.0002 | 0.03 | 0.015 |
| | JM-1L-DI-wide | -0.0009 | -1.70 | 0.026 | -0.0013 | -0.28 | 0.026 | -0.0002 | -0.03 | 0.015 |
| | FCS-1L-DI-wide | -0.0009 | -1.71 | 0.026 | -0.0013 | -0.28 | 0.026 | -0.0001 | -0.03 | 0.015 |
| 10 clusters | JM-2L-wide | -0.0009 | -1.71 | 0.026 | -0.0013 | -0.28 | 0.026 | -0.0001 | -0.03 | 0.015 |
| | FCS-2L-wide | -0.0008 | -1.70 | 0.026 | -0.0012 | -0.28 | 0.026 | -0.0002 | -0.03 | 0.015 |
| | SMC-JM-2L-DI | -0.0009 | -1.71 | 0.026 | -0.0013 | -0.28 | 0.026 | -0.0001 | -0.03 | 0.015 |
| | SMC-SM-2L-DI | -0.0009 | -1.75 | 0.026 | -0.0012 | -0.27 | 0.026 | -0.0001 | -0.02 | 0.015 |
| | SMC-JM-3L | -0.0009 | -1.72 | 0.026 | -0.0012 | -0.28 | 0.026 | -0.0001 | -0.02 | 0.015 |

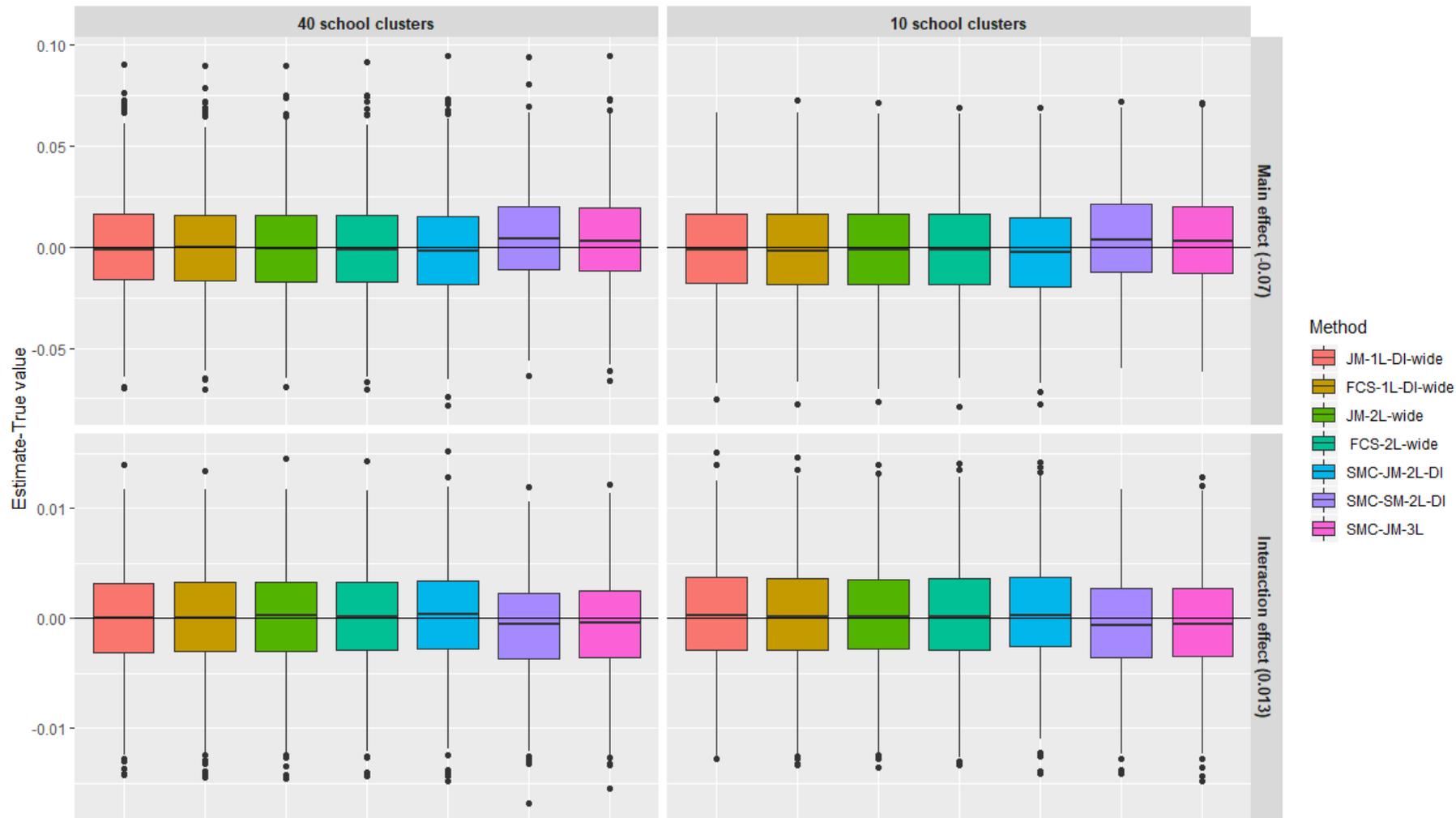

Figure S1: Distribution of the bias in the estimated regression coefficient for the main effect ($\beta_1$, $true\ value = -0.07$) and the interaction effect ($\beta_3$, $true\ value = 0.013$) in analysis model 1, across the 1000 simulated datasets for the 7 multiple imputation (MI) approaches under two scenarios for number of higher level clusters (40 school clusters and 10 school clusters) when data are missing at random with inflated dependencies (MAR-inflated)

The lower and upper margins of the boxes represent the 25$^{th}$ (Q$_1$) and the 75$^{th}$ (Q$_3$) percentiles of the distribution respectively. The whiskers extend to Q$_1$-1.5*(Q$_3$- Q$_1$) at the bottom and Q$_3$ +1.5*(Q$_3$- Q$_1$) at the top.

The following abbreviations are used to denote different MI methods, e.g., DI: dummy indicators, FCS: fully conditional specification, JM: joint modelling, SM: sequential modelling, SMC: substantive model compatible.

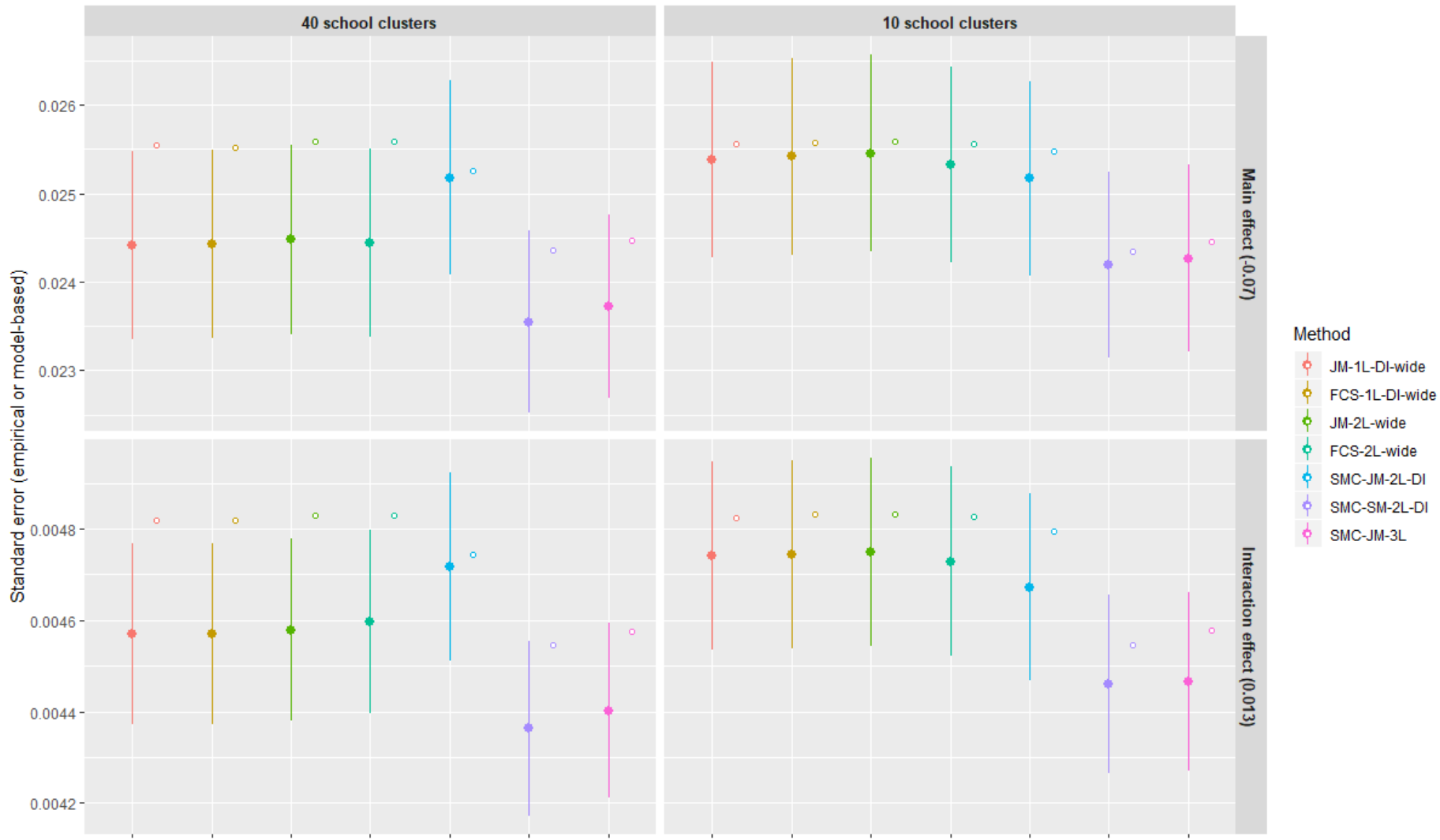

Figure S2: Empirical standard errors (filled circles with error bars showing ±1.96× Monte Carlo standard errors) and average model-based standard errors (hollow circles) in analysis model 1 from 1000 simulated datasets, for the 7 multiple imputation (MI) approaches under two scenarios for number of higher level clusters (40 school clusters and 10 school clusters) when data are missing at random with inflated dependencies (MAR-inflated)
The following abbreviations are used to denote different MI methods, e.g., DI: dummy indicators, FCS: fully conditional specification, JM: joint modelling, SM: sequential modelling, SMC: substantive model compatible.

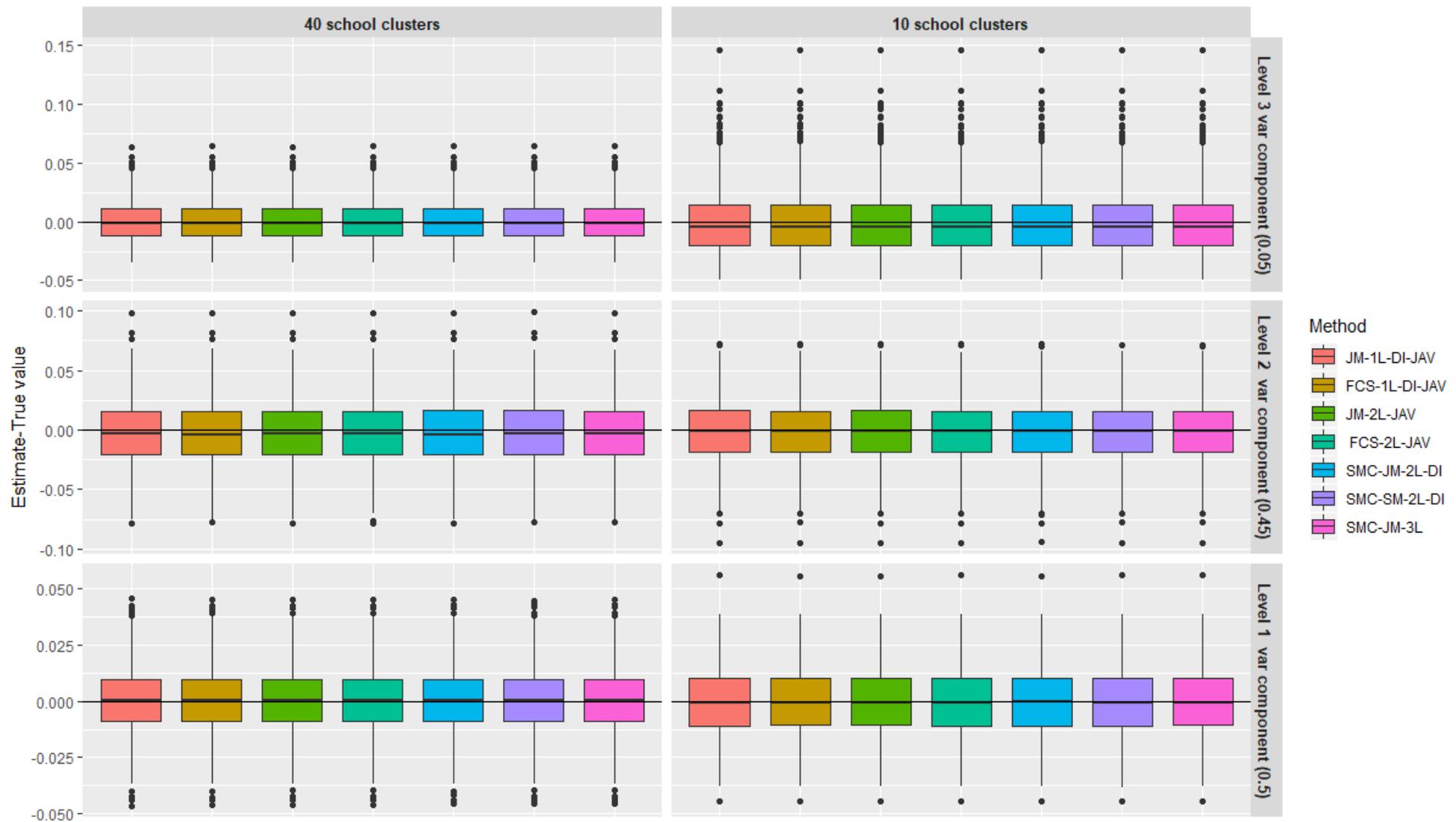

Figure S3 Distribution of the bias in the estimated variance components (VC) at level 1, 2 and 3 in analysis model 1 across the 1000 simulated datasets for the 7 multiple imputation (MI) approaches under two scenarios for number of higher level clusters (40 school clusters and 10 school clusters) when data are missing at random with inflated dependencies (MAR-inflated) The following abbreviations are used to denote different MI methods, e.g., DI: dummy indicators, FCS: fully conditional specification, JM: joint modelling, SM: sequential modelling, SMC: substantive model compatible.

# Supplement C: Performance of multiple imputation (MI) methods under analysis model 2

Table S6: Performance of the 11 multiple imputation (MI) methods for estimating the main effect ($\beta_1 = -0.024$) and the interaction effect ($\beta_3 = 0.023$) in the analysis model 2 when data are missing at random with dependencies based on the CATS data (MAR-CATS)

| No of higher-level clusters | Method | Main effect (-0.024) | | | | | | Interaction effect (0.023) | | | | | |
|---|---|---|---|---|---|---|---|---|---|---|---|---|---|
| | | Average Estimate | Bias | Relative Bias (%) | Emp SE | Model SE | Coverage | Average Estimate | Bias | Relative Bias (%) | Emp SE | Model SE | Coverage |
| 40 clusters | JM-1L-DI-wide | -0.024 | -0.0001 | 0.4 | 0.0094 | 0.0092 | 95.0% | 0.018 | -0.0048 | -21.0 | 0.0076 | 0.0092 | 95.9% |
| | JM-1L-DI-wide-JAV | -0.023 | 0.0009 | -3.6 | 0.0095 | 0.0093 | 94.5% | 0.021 | -0.0015 | -6.6 | 0.0113 | 0.0115 | 94.8% |
| | FCS-1L-DI-wide | -0.024 | -0.0001 | 0.3 | 0.0094 | 0.0092 | 94.0% | 0.018 | -0.0048 | -21.0 | 0.0076 | 0.0092 | 95.7% |
| | FCS-1L-DI-wide-passive_c | -0.024 | 0.0001 | -0.3 | 0.0094 | 0.0092 | 94.8% | 0.020 | -0.0031 | -13.3 | 0.0084 | 0.0094 | 96.0% |
| | FCS-1L-DI-wide-passive_all | -0.024 | 0.0002 | -0.9 | 0.0094 | 0.0092 | 94.1% | 0.021 | -0.0021 | -9.3 | 0.0089 | 0.0095 | 96.0% |
| | JM-2L-wide-JAV | -0.023 | 0.0007 | -2.8 | 0.0095 | 0.0093 | 94.5% | 0.022 | -0.0010 | -4.3 | 0.0116 | 0.0116 | 95.1% |
| | FCS-2L-wide-passive_c | -0.024 | -0.0001 | 0.5 | 0.0095 | 0.0093 | 94.3% | 0.020 | -0.0028 | -12.4 | 0.0085 | 0.0094 | 95.9% |
| | FCS-2L-wide-passive_all | -0.024 | 0.0001 | -0.3 | 0.0094 | 0.0093 | 94.3% | 0.021 | -0.0019 | -8.2 | 0.0090 | 0.0095 | 95.6% |
| | SMC-JM-2L-DI | -0.024 | -0.0003 | 1.4 | 0.0095 | 0.0091 | 94.1% | 0.023 | -0.0003 | -1.3 | 0.0097 | 0.0093 | 93.3% |
| | SMC-SM-2L-DI | -0.023 | 0.0007 | -2.7 | 0.0095 | 0.0093 | 95.0% | 0.023 | 0.0000 | 0.0 | 0.0095 | 0.0096 | 94.7% |
| | SMC-JM-3L | -0.024 | 0.0005 | -2.0 | 0.0096 | 0.0093 | 95.0% | 0.023 | 0.0003 | 1.1 | 0.0096 | 0.0096 | 94.5% |
| 10 clusters | JM-1L-DI-wide | -0.024 | 0.0002 | -0.7 | 0.0093 | 0.0092 | 95.3% | 0.018 | -0.0048 | -21.0 | 0.0077 | 0.0092 | 95.0% |
| | JM-1L-DI-wide-JAV | -0.023 | 0.0011 | -4.5 | 0.0093 | 0.0093 | 94.8% | 0.021 | -0.0021 | -9.1 | 0.0114 | 0.0116 | 94.7% |
| | FCS-1L-DI-wide | -0.024 | 0.0002 | -0.9 | 0.0092 | 0.0092 | 95.4% | 0.018 | -0.0049 | -21.1 | 0.0077 | 0.0092 | 95.7% |
| | FCS-1L-DI-wide-passive_c | -0.024 | 0.0003 | -1.3 | 0.0092 | 0.0093 | 95.3% | 0.020 | -0.0030 | -13.1 | 0.0084 | 0.0094 | 95.9% |
| | FCS-1L-DI-wide-passive_all | -0.024 | 0.0005 | -2.1 | 0.0092 | 0.0092 | 94.9% | 0.021 | -0.0021 | -9.1 | 0.0089 | 0.0095 | 96.1% |
| | JM-2L-wide-JAV | -0.023 | 0.0011 | -4.5 | 0.0092 | 0.0093 | 95.0% | 0.021 | -0.0021 | -9.1 | 0.0115 | 0.0116 | 94.4% |
| | FCS-2L-wide-passive_c | -0.024 | 0.0003 | -1.4 | 0.0092 | 0.0093 | 95.2% | 0.020 | -0.0031 | -13.3 | 0.0084 | 0.0094 | 96.2% |
| | FCS-2L-wide-passive_all | -0.024 | 0.0005 | -2.0 | 0.0091 | 0.0093 | 95.4% | 0.021 | -0.0021 | -9.1 | 0.0089 | 0.0095 | 96.2% |
| | SMC-JM-2L-DI | -0.024 | 0.0001 | -0.2 | 0.0093 | 0.0091 | 95.0% | 0.023 | -0.0004 | -1.6 | 0.0096 | 0.0093 | 93.9% |
| | SMC-SM-2L-DI | -0.023 | 0.0010 | -4.2 | 0.0093 | 0.0093 | 95.3% | 0.023 | -0.0003 | -1.3 | 0.0096 | 0.0095 | 95.4% |
| | SMC-JM-3L | -0.023 | 0.0008 | -3.2 | 0.0093 | 0.0093 | 95.4% | 0.023 | -0.0002 | -0.7 | 0.0097 | 0.0096 | 95.4% |

Table S7: Performance of the 11 multiple imputation (MI) methods in estimating the variance components (VC) at level 3, 2 and 1 in the analysis model 2 when data are missing at random with dependencies based on the CATS data (MAR-CATS)

| No of higher-level clusters | Method | Level 3 VC (0.05) | | | Level 2 VC (0.45) | | | Level 1 VC (0.5) | | |
|---|---|---|---|---|---|---|---|---|---|---|
| | | Bias | Relative Bias (%) | Emp SE | Bias | Relative Bias (%) | Emp SE | Bias | Relative Bias (%) | Emp SE |
| 40 clusters | JM-1L-DI-wide | 0.0000 | 0.04 | 0.016 | 0.0000 | -0.01 | 0.026 | 0.0000 | 0.00 | 0.014 |
| | JM-1L-DI-wide-JAV | 0.0001 | 0.10 | 0.016 | -0.0001 | -0.03 | 0.026 | -0.0004 | -0.07 | 0.014 |
| | FCS-1L-DI-wide | 0.0000 | 0.05 | 0.016 | -0.0001 | -0.01 | 0.026 | 0.0000 | 0.00 | 0.014 |
| | FCS-1L-DI-wide-passive_c | 0.0000 | -0.01 | 0.016 | -0.0003 | -0.07 | 0.026 | -0.0001 | -0.02 | 0.014 |
| | FCS-1L-DI-wide-passive_all | 0.0000 | 0.00 | 0.016 | -0.0002 | -0.04 | 0.026 | -0.0003 | -0.06 | 0.014 |
| | JM-2L-wide-JAV | 0.0000 | -0.05 | 0.016 | 0.0000 | -0.01 | 0.026 | -0.0004 | -0.08 | 0.014 |
| | FCS-2L-wide-passive_c | 0.0000 | -0.05 | 0.016 | -0.0003 | -0.07 | 0.026 | -0.0001 | -0.03 | 0.014 |
| | FCS-2L-wide-passive_all | 0.0000 | -0.02 | 0.016 | -0.0001 | -0.03 | 0.026 | -0.0003 | -0.07 | 0.014 |
| | SMC-JM-2L-DI | 0.0000 | 0.07 | 0.016 | -0.0002 | -0.05 | 0.026 | -0.0005 | -0.10 | 0.014 |
| | SMC-SM-2L-DI | -0.0001 | -0.12 | 0.016 | 0.0001 | 0.01 | 0.026 | -0.0005 | -0.10 | 0.014 |
| | SMC-JM-3L | 0.0000 | 0.01 | 0.016 | -0.0001 | -0.02 | 0.026 | -0.0005 | -0.10 | 0.014 |
| 10 clusters | JM-1L-DI-wide | -0.0003 | -0.59 | 0.025 | 0.0006 | 0.14 | 0.026 | 0.0005 | 0.10 | 0.014 |
| | JM-1L-DI-wide-JAV | -0.0003 | -0.57 | 0.025 | 0.0006 | 0.14 | 0.026 | 0.0002 | 0.05 | 0.014 |
| | FCS-1L-DI-wide | -0.0003 | -0.57 | 0.025 | 0.0007 | 0.15 | 0.026 | 0.0005 | 0.11 | 0.014 |
| | FCS-1L-DI-wide-passive_c | -0.0003 | -0.64 | 0.025 | 0.0004 | 0.08 | 0.026 | 0.0004 | 0.09 | 0.014 |
| | FCS-1L-DI-wide-passive_all | -0.0003 | -0.64 | 0.025 | 0.0005 | 0.12 | 0.026 | 0.0003 | 0.05 | 0.014 |
| | JM-2L-wide-JAV | -0.0003 | -0.61 | 0.025 | 0.0006 | 0.14 | 0.026 | 0.0002 | 0.05 | 0.014 |
| | FCS-2L-wide-passive_c | -0.0003 | -0.66 | 0.025 | 0.0004 | 0.09 | 0.026 | 0.0004 | 0.09 | 0.014 |
| | FCS-2L-wide-passive_all | -0.0003 | -0.63 | 0.025 | 0.0006 | 0.12 | 0.026 | 0.0002 | 0.05 | 0.014 |
| | SMC-JM-2L-DI | -0.0003 | -0.60 | 0.025 | 0.0005 | 0.10 | 0.026 | 0.0001 | 0.01 | 0.014 |
| | SMC-SM-2L-DI | -0.0004 | -0.71 | 0.025 | 0.0007 | 0.16 | 0.026 | 0.0001 | 0.01 | 0.014 |
| | SMC-JM-3L | -0.0003 | -0.60 | 0.025 | 0.0006 | 0.14 | 0.026 | 0.0001 | 0.01 | 0.014 |

Table S8: Performance of the 11 multiple imputation (MI) methods for estimating the main effect ($\beta_1 = -0.024$) and the interaction effect ($\beta_3 = 0.023$) in the analysis model 2 when data are missing at random with inflated dependencies (MAR-inflated)

| No of higher-level clusters | Method | Main effect (-0.024) | | | | | | Interaction effect (0.023) | | | | | |
|---|---|---|---|---|---|---|---|---|---|---|---|---|---|
| | | Average Estimate | Bias | Relative Bias (%) | Emp SE | Model SE | Coverage | Average Estimate | Bias | Relative Bias (%) | Emp SE | Model SE | Coverage |
| 40 clusters | JM-1L-DI-wide | -0.025 | -0.0005 | 2.2 | 0.0088 | 0.0093 | 95.6% | 0.018 | -0.0049 | -21.3 | 0.0075 | 0.0092 | 95.5% |
| | JM-1L-DI-wide-JAV | -0.024 | 0.0005 | -1.9 | 0.0088 | 0.0094 | 95.6% | 0.021 | -0.0025 | -10.8 | 0.0119 | 0.0118 | 93.2% |
| | FCS-1L-DI-wide | -0.025 | -0.0006 | 2.4 | 0.0088 | 0.0093 | 95.8% | 0.018 | -0.0049 | -21.3 | 0.0076 | 0.0092 | 95.8% |
| | FCS-1L-DI-wide-passive_c | -0.024 | -0.0004 | 1.6 | 0.0088 | 0.0093 | 95.9% | 0.020 | -0.0031 | -13.3 | 0.0084 | 0.0094 | 95.9% |
| | FCS-1L-DI-wide-passive_all | -0.024 | -0.0002 | 0.8 | 0.0088 | 0.0094 | 95.8% | 0.021 | -0.0021 | -9.0 | 0.0090 | 0.0096 | 95.8% |
| | JM-2L-wide-JAV | -0.024 | 0.0003 | -1.1 | 0.0089 | 0.0094 | 95.8% | 0.021 | -0.0019 | -8.2 | 0.0121 | 0.0119 | 93.7% |
| | FCS-2L-wide-passive_c | -0.024 | -0.0005 | 2.0 | 0.0088 | 0.0094 | 95.6% | 0.020 | -0.0028 | -12.4 | 0.0085 | 0.0094 | 96.1% |
| | FCS-2L-wide-passive_all | -0.024 | -0.0004 | 1.6 | 0.0088 | 0.0094 | 95.8% | 0.021 | -0.0018 | -7.8 | 0.0090 | 0.0096 | 95.4% |
| | SMC-JM-2L-DI | -0.025 | -0.0010 | 4.0 | 0.0090 | 0.0092 | 95.0% | 0.023 | 0.0000 | -0.2 | 0.0099 | 0.0093 | 92.8% |
| | SMC-SM-2L-DI | -0.023 | 0.0007 | -3.0 | 0.0088 | 0.0094 | 96.3% | 0.023 | 0.0002 | 0.7 | 0.0097 | 0.0096 | 94.8% |
| | SMC-JM-3L | -0.024 | 0.0005 | -1.9 | 0.0089 | 0.0094 | 96.0% | 0.023 | 0.0003 | 1.5 | 0.0098 | 0.0096 | 94.5% |
| 10 clusters | JM-1L-DI-wide | -0.024 | -0.0001 | 0.5 | 0.0092 | 0.0093 | 95.3% | 0.018 | -0.0051 | -22.1 | 0.0075 | 0.0093 | 95.6% |
| | JM-1L-DI-wide-JAV | -0.023 | 0.0008 | -3.3 | 0.0092 | 0.0094 | 94.8% | 0.021 | -0.0020 | -8.5 | 0.0122 | 0.0119 | 93.9% |
| | FCS-1L-DI-wide | -0.024 | -0.0002 | 0.8 | 0.0092 | 0.0093 | 95.2% | 0.018 | -0.0051 | -22.0 | 0.0076 | 0.0093 | 95.3% |
| | FCS-1L-DI-wide-passive_c | -0.024 | 0.0000 | 0.0 | 0.0092 | 0.0093 | 95.1% | 0.020 | -0.0032 | -14.0 | 0.0084 | 0.0094 | 96.0% |
| | FCS-1L-DI-wide-passive_all | -0.024 | 0.0001 | -0.6 | 0.0092 | 0.0093 | 95.0% | 0.021 | -0.0021 | -9.2 | 0.0090 | 0.0096 | 95.7% |
| | JM-2L-wide-JAV | -0.023 | 0.0008 | -3.3 | 0.0092 | 0.0093 | 94.7% | 0.021 | -0.0020 | -8.5 | 0.0122 | 0.0119 | 93.4% |
| | FCS-2L-wide-passive_c | -0.024 | 0.0000 | 0.0 | 0.0092 | 0.0093 | 95.3% | 0.020 | -0.0032 | -14.0 | 0.0084 | 0.0094 | 95.7% |
| | FCS-2L-wide-passive_all | -0.024 | 0.0002 | -0.7 | 0.0092 | 0.0093 | 95.1% | 0.021 | -0.0022 | -9.4 | 0.0091 | 0.0096 | 94.8% |
| | SMC-JM-2L-DI | -0.024 | -0.0004 | 1.8 | 0.0093 | 0.0092 | 94.9% | 0.022 | -0.0005 | -2.2 | 0.0098 | 0.0093 | 92.8% |
| | SMC-SM-2L-DI | -0.023 | 0.0012 | -4.8 | 0.0092 | 0.0094 | 95.1% | 0.023 | -0.0004 | -1.7 | 0.0095 | 0.0095 | 94.7% |
| | SMC-JM-3L | -0.023 | 0.0009 | -3.7 | 0.0093 | 0.0094 | 94.3% | 0.023 | -0.0001 | -0.6 | 0.0097 | 0.0096 | 94.0% |

Table S9: Performance of the 11 multiple imputation (MI) methods in estimating the variance components (VC) at level 3, 2 and 1 in the analysis model 2 when data are missing at random with inflated dependencies (MAR-inflated)

| No of higher-level clusters | Method | Level 3 VC (0.05) | Level 2 VC (0.45) | Level 1 VC (0.5) |
|---|---|---|---|---|

|  |  | Bias | Relative Bias (%) | Emp SE | Bias | Relative Bias (%) | Emp SE | Bias | Relative Bias (%) | Emp SE |
|---|---|---|---|---|---|---|---|---|---|---|
| 40 clusters | JM-1L-DI-wide | 0.0000 | 0.09 | 0.016 | 0.0010 | 0.22 | 0.026 | 0.0005 | 0.09 | 0.014 |
|  | JM-1L-DI-wide-JAV | 0.0001 | 0.16 | 0.016 | 0.0009 | 0.20 | 0.026 | 0.0002 | 0.04 | 0.014 |
|  | FCS-1L-DI-wide | 0.0000 | 0.09 | 0.016 | 0.0010 | 0.21 | 0.026 | 0.0005 | 0.09 | 0.014 |
|  | FCS-1L-DI-wide-passive_c | 0.0000 | 0.04 | 0.016 | 0.0007 | 0.15 | 0.026 | 0.0004 | 0.07 | 0.014 |
|  | FCS-1L-DI-wide-passive_all | 0.0000 | 0.05 | 0.016 | 0.0009 | 0.19 | 0.026 | 0.0002 | 0.03 | 0.014 |
|  | JM-2L-wide-JAV | 0.0000 | 0.02 | 0.016 | 0.0010 | 0.21 | 0.026 | 0.0002 | 0.03 | 0.014 |
|  | FCS-2L-wide-passive_c | 0.0000 | -0.01 | 0.016 | 0.0007 | 0.15 | 0.026 | 0.0004 | 0.07 | 0.014 |
|  | FCS-2L-wide-passive_all | 0.0000 | 0.01 | 0.016 | 0.0009 | 0.19 | 0.026 | 0.0002 | 0.03 | 0.014 |
|  | SMC-JM-2L-DI | 0.0001 | 0.13 | 0.016 | 0.0007 | 0.16 | 0.026 | -0.0001 | -0.01 | 0.014 |
|  | SMC-SM-2L-DI | 0.0000 | -0.04 | 0.016 | 0.0012 | 0.26 | 0.026 | 0.0000 | 0.00 | 0.014 |
|  | SMC-JM-3L | 0.0000 | 0.09 | 0.016 | 0.0010 | 0.23 | 0.026 | 0.0000 | 0.00 | 0.014 |
| 10 clusters | JM-1L-DI-wide | -0.0006 | -1.10 | 0.027 | 0.0000 | -0.01 | 0.026 | 0.0000 | 0.00 | 0.014 |
|  | JM-1L-DI-wide-JAV | -0.0005 | -1.05 | 0.027 | 0.0000 | 0.00 | 0.026 | -0.0003 | -0.07 | 0.015 |
|  | FCS-1L-DI-wide | -0.0006 | -1.13 | 0.027 | 0.0000 | -0.01 | 0.026 | 0.0000 | 0.00 | 0.015 |
|  | FCS-1L-DI-wide-passive_c | -0.0006 | -1.20 | 0.027 | -0.0003 | -0.07 | 0.026 | -0.0001 | -0.02 | 0.014 |
|  | FCS-1L-DI-wide-passive_all | -0.0006 | -1.12 | 0.027 | -0.0001 | -0.03 | 0.026 | -0.0003 | -0.06 | 0.015 |
|  | JM-2L-wide-JAV | -0.0005 | -1.09 | 0.027 | 0.0000 | 0.01 | 0.026 | -0.0004 | -0.07 | 0.015 |
|  | FCS-2L-wide-passive_c | -0.0006 | -1.20 | 0.027 | -0.0003 | -0.07 | 0.026 | -0.0001 | -0.02 | 0.015 |
|  | FCS-2L-wide-passive_all | -0.0006 | -1.17 | 0.027 | -0.0001 | -0.03 | 0.026 | -0.0003 | -0.06 | 0.015 |
|  | SMC-JM-2L-DI | -0.0006 | -1.13 | 0.027 | -0.0002 | -0.05 | 0.026 | -0.0005 | -0.10 | 0.015 |
|  | SMC-SM-2L-DI | -0.0006 | -1.22 | 0.027 | 0.0001 | 0.03 | 0.026 | -0.0005 | -0.09 | 0.015 |
|  | SMC-JM-3L | -0.0005 | -1.10 | 0.027 | 0.0000 | 0.00 | 0.026 | -0.0005 | -0.09 | 0.015 |

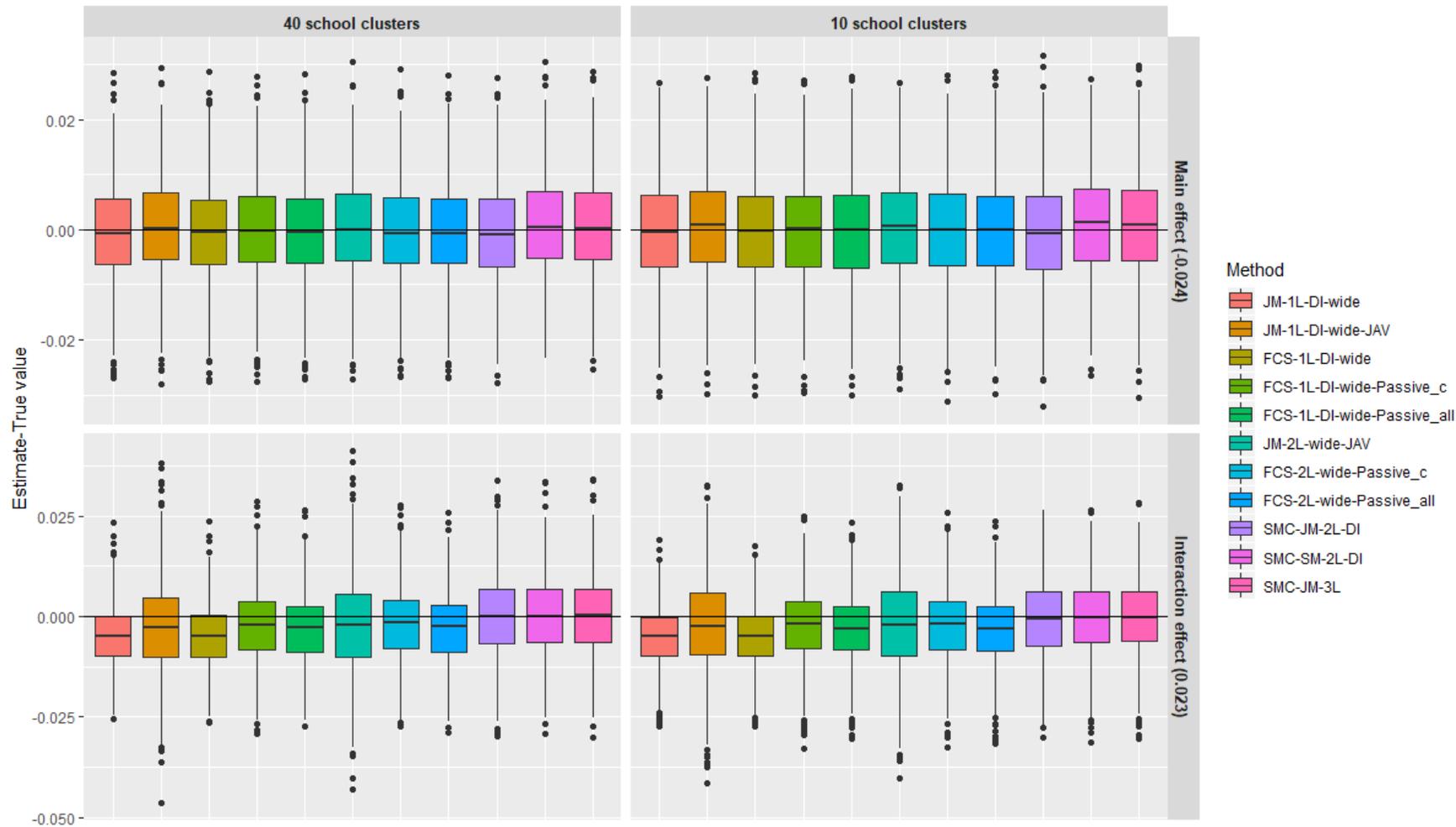

Figure S4: Distribution of the bias in the estimated regression coefficient for the main effect ($\beta_1$, $true\ value = -0.024$) and the interaction effect ($\beta_3$, $true\ value = 0.023$) in analysis model 2, across the 1000 simulated datasets for the 11 multiple imputation (MI) approaches under two scenarios for number of higher level clusters( 40 school clusters and 10 school clusters) when data are missing at random with inflated dependencies (MAR-inflated)

The lower and upper margins of the boxes represent the $25^{th}$ ($Q_1$) and the $75^{th}$ ($Q_3$) percentiles of the distribution respectively. The whiskers extend to $Q_1-1.5*(Q_3-Q_1)$ at the bottom and $Q_3+1.5*(Q_3-Q_1)$ at the top.

The following abbreviations are used to denote different MI methods, e.g., DI: dummy indicators, FCS: fully conditional specification, JM: joint modelling, JAV: Just another variable, passive_all: passive imputation (with all two-way interactions between the NAPLAN scores at each of the 3 waves and SES as predictors in imputing incomplete depressive symptoms),

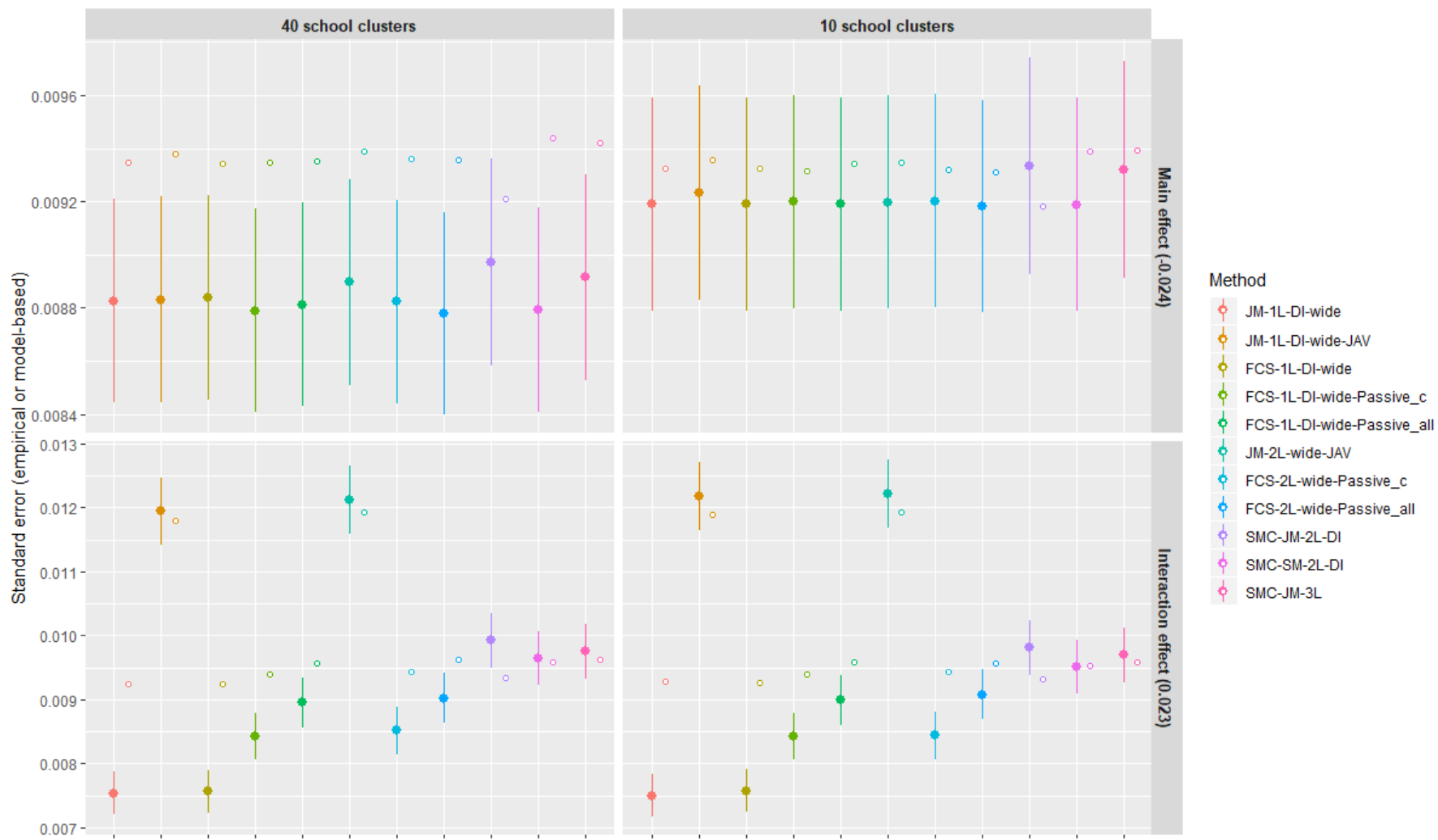

Figure S5: Empirical standard errors (filled circles with error bars showing ±1.96× Monte Carlo standard errors) and average model-based standard errors (hollow circles) in analysis model 2 from 1000 simulated datasets, for the 11 multiple imputation (MI) approaches under two scenarios for number of higher level clusters (40 school clusters and 10 school clusters) when data are missing at random with inflated dependencies (MAR-Inflated)

The following abbreviations are used to denote different MI methods, e.g., DI: dummy indicators, FCS: fully conditional specification, JM: joint modelling, JAV: Just another variable, passive_all: passive imputation (with all two-way interactions between the NAPLAN scores at each of the 3 waves and SES as predictors in imputing incomplete depressive symptoms),

passive_c: passive imputation (with single interaction between the NAPLAN score at the next wave and SES as a predictor in imputing incomplete depressive symptoms), SM: sequential modelling, SMC: substantive model compatible.

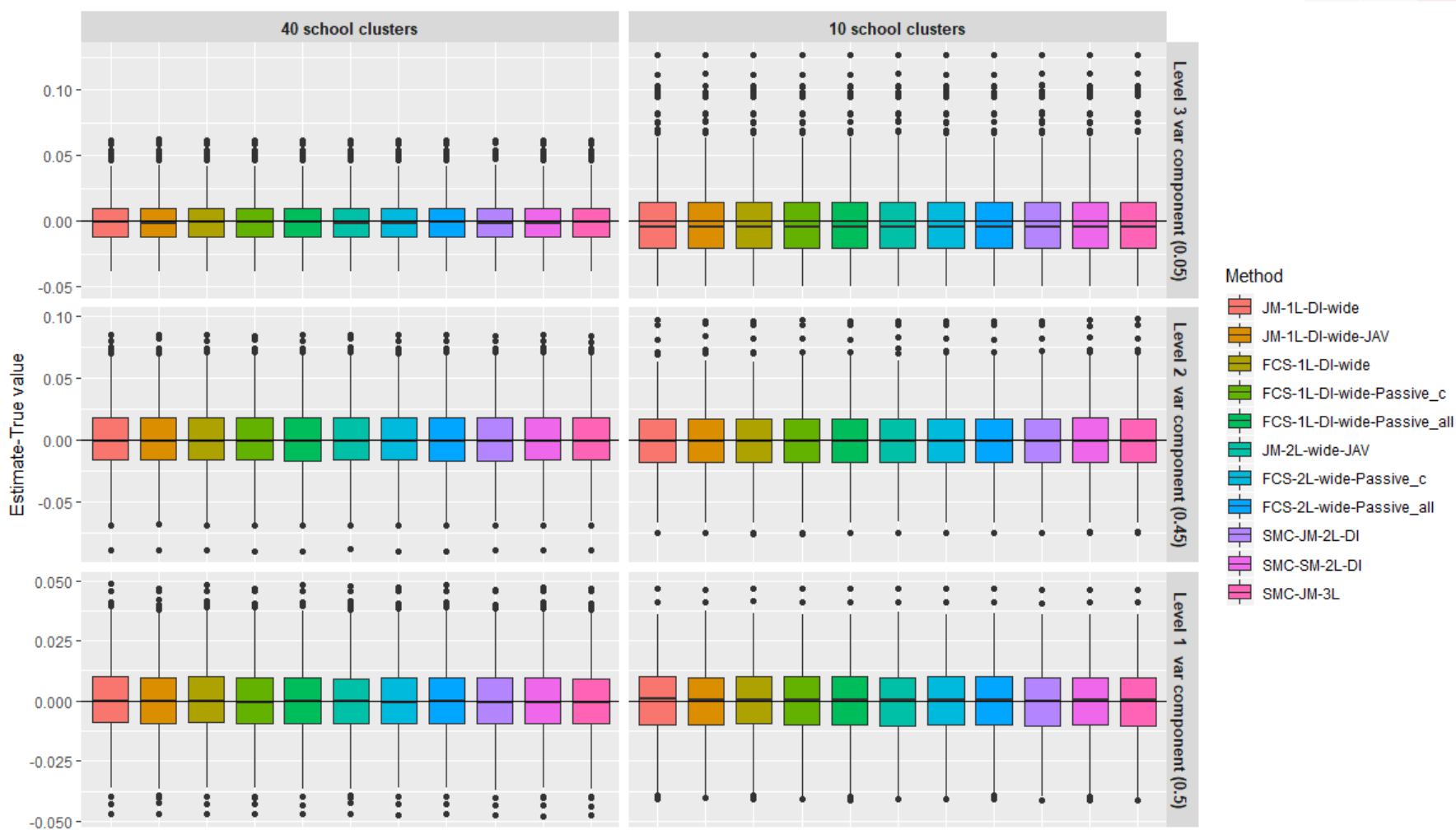

Figure S6: Distribution of the bias in the estimated variance components (VC) at level 1, 2 and 3 in analysis model 2 across the 1000 simulated datasets for the 11 multiple imputation (MI) approaches under two scenarios for number of higher level clusters (40 school clusters and 10 school clusters) when data are missing at random with inflated dependencies (MAR-inflated) The following abbreviations are used to denote different MI methods, e.g., DI: dummy indicators, FCS: fully conditional specification, JM: joint modelling, JAV: Just another variable, passive_all: passive imputation (with all two-way interactions between the NAPLAN scores at each of the 3 waves and SES as predictors in imputing incomplete depressive symptoms),

passive_c: passive imputation (with single interaction between the NAPLAN score at the next wave and SES as a predictor in imputing incomplete depressive symptoms), SM: sequential modelling, SMC: substantive model compatible

## Supplement D: Performance of multiple imputation (MI) methods under analysis model 3

Table S10: Performance of the 9 multiple imputation (MI) methods for estimating the main effect ($\beta_1 = -0.024$) and the quadratic effect ($\beta_3 = -0.009$) in the analysis model 3 when data are missing at random with dependencies based on the CATS data (MAR-CATS)

| No of higher-level clusters | Method | Main effect (-0.024) | | | | | | Quadratic effect (-0.009) | | | | | |
|---|---|---|---|---|---|---|---|---|---|---|---|---|---|
| | | Average Estimate | Bias | Relative Bias (%) | Emp SE | Model SE | Coverage | Average Estimate | Bias | Relative Bias (%) | Emp SE | Model SE | Coverage |
| 40 clusters | JM-1L-DI-wide | -0.028 | -0.0037 | 15.4 | 0.0091 | 0.0092 | 93.3% | -0.007 | 0.0023 | -25.5 | 0.0026 | 0.0035 | 95.9% |
| | JM-1L-DI-wide-JAV | -0.020 | 0.0041 | -16.9 | 0.0105 | 0.0107 | 92.9% | -0.006 | 0.0027 | -29.8 | 0.0044 | 0.0044 | 90.6% |
| | FCS-1L-DI-wide | -0.028 | -0.0038 | 15.7 | 0.0091 | 0.0092 | 93.7% | -0.007 | 0.0023 | -25.7 | 0.0026 | 0.0035 | 95.4% |
| | FCS-1L-DI-wide-passive | -0.028 | -0.0037 | 15.6 | 0.0091 | 0.0092 | 93.8% | -0.007 | 0.0023 | -26.0 | 0.0026 | 0.0035 | 95.1% |
| | JM-2L-wide-JAV | -0.020 | 0.0041 | -16.9 | 0.0105 | 0.0107 | 92.4% | -0.006 | 0.0025 | -28.3 | 0.0044 | 0.0045 | 91.4% |
| | FCS-2L-wide-passive | -0.028 | -0.0037 | 15.3 | 0.0091 | 0.0092 | 93.2% | -0.007 | 0.0023 | -25.9 | 0.0026 | 0.0035 | 95.1% |
| | SMC-JM-2L-DI | -0.024 | 0.0000 | 0.0 | 0.0094 | 0.0092 | 93.9% | -0.009 | -0.0001 | 1.1 | 0.0037 | 0.0036 | 93.8% |
| | SMC-SM-2L-DI | -0.024 | 0.0003 | -1.2 | 0.0089 | 0.0093 | 95.4% | -0.009 | -0.0001 | 1.0 | 0.0034 | 0.0035 | 95.3% |
| | SMC-JM-3L | -0.024 | 0.0005 | -2.1 | 0.0092 | 0.0095 | 94.9% | -0.009 | -0.0001 | 1.5 | 0.0036 | 0.0036 | 94.5% |
| 10 clusters | JM-1L-DI-wide | -0.028 | -0.0038 | 15.9 | 0.0092 | 0.0093 | 93.4% | -0.007 | 0.0024 | -26.6 | 0.0026 | 0.0035 | 94.6% |
| | JM-1L-DI-wide-JAV | -0.020 | 0.0037 | -15.6 | 0.0105 | 0.0107 | 93.8% | -0.006 | 0.0027 | -30.1 | 0.0044 | 0.0045 | 90.4% |
| | FCS-1L-DI-wide | -0.028 | -0.0038 | 16.0 | 0.0093 | 0.0092 | 93.3% | -0.007 | 0.0024 | -26.4 | 0.0026 | 0.0035 | 95.4% |
| | FCS-1L-DI-wide-passive | -0.028 | -0.0039 | 16.3 | 0.0092 | 0.0092 | 92.8% | -0.007 | 0.0024 | -27.2 | 0.0026 | 0.0035 | 94.5% |
| | JM-2L-wide-JAV | -0.020 | 0.0037 | -15.4 | 0.0105 | 0.0107 | 94.0% | -0.006 | 0.0027 | -30.1 | 0.0043 | 0.0045 | 91.2% |
| | FCS-2L-wide-passive | -0.028 | -0.0039 | 16.2 | 0.0092 | 0.0092 | 92.7% | -0.007 | 0.0024 | -27.0 | 0.0026 | 0.0035 | 94.6% |
| | SMC-JM-2L-DI | -0.024 | -0.0001 | 0.4 | 0.0095 | 0.0092 | 94.8% | -0.009 | 0.0000 | 0.0 | 0.0037 | 0.0036 | 94.9% |
| | SMC-SM-2L-DI | -0.024 | -0.0001 | 0.4 | 0.0092 | 0.0093 | 95.2% | -0.009 | 0.0001 | -1.0 | 0.0034 | 0.0035 | 95.6% |
| | SMC-JM-3L | -0.024 | 0.0001 | -0.6 | 0.0095 | 0.0094 | 95.2% | -0.009 | 0.0000 | -0.5 | 0.0036 | 0.0036 | 94.9% |

Table S11: Performance of the 9 multiple imputation (MI) methods in estimating the variance components (VC) at level 3, 2 and 1 in the analysis model 3 when data are missing at random with dependencies based on the CATS data (MAR-CATS)

| No of higher-level clusters | Method | Level 3 VC (0.05) | | | Level 2 VC (0.45) | | | Level 1 VC (0.5) | | |
|---|---|---|---|---|---|---|---|---|---|---|
| | | Bias | Relative Bias (%) | Emp SE | Bias | Relative Bias (%) | Emp SE | Bias | Relative Bias (%) | Emp SE |
| 40 clusters | JM-1L-DI-wide | 0.0000 | 0.06 | 0.016 | -0.0004 | -0.08 | 0.026 | 0.0003 | 0.07 | 0.015 |
| | JM-1L-DI-wide-JAV | 0.0001 | 0.17 | 0.016 | -0.0001 | -0.03 | 0.026 | 0.0005 | 0.10 | 0.015 |
| | FCS-1L-DI-wide | 0.0000 | 0.06 | 0.016 | -0.0004 | -0.08 | 0.026 | 0.0003 | 0.07 | 0.015 |
| | FCS-1L-DI-wide-passive | 0.0000 | 0.06 | 0.016 | -0.0004 | -0.08 | 0.026 | 0.0004 | 0.07 | 0.015 |
| | JM-2L-wide-JAV | 0.0000 | 0.05 | 0.016 | -0.0001 | -0.02 | 0.026 | 0.0005 | 0.09 | 0.015 |
| | FCS-2L-wide-passive | 0.0000 | 0.02 | 0.016 | -0.0003 | -0.08 | 0.026 | 0.0004 | 0.07 | 0.015 |
| | SMC-JM-2L-DI | 0.0000 | 0.05 | 0.016 | -0.0004 | -0.10 | 0.026 | 0.0000 | 0.00 | 0.014 |
| | SMC-SM-2L-DI | 0.0000 | -0.07 | 0.016 | -0.0004 | -0.09 | 0.026 | 0.0000 | 0.00 | 0.015 |
| | SMC-JM-3L | 0.0000 | 0.03 | 0.016 | -0.0005 | -0.11 | 0.026 | 0.0000 | 0.00 | 0.015 |
| 10 clusters | JM-1L-DI-wide | -0.0017 | -3.41 | 0.027 | -0.0012 | -0.27 | 0.027 | -0.0002 | -0.03 | 0.014 |
| | JM-1L-DI-wide-JAV | -0.0017 | -3.31 | 0.027 | -0.0009 | -0.21 | 0.027 | -0.0001 | -0.02 | 0.014 |
| | FCS-1L-DI-wide | -0.0017 | -3.40 | 0.027 | -0.0012 | -0.27 | 0.027 | -0.0002 | -0.03 | 0.014 |
| | FCS-1L-DI-wide-passive | -0.0017 | -3.40 | 0.027 | -0.0012 | -0.27 | 0.027 | -0.0002 | -0.03 | 0.014 |
| | JM-2L-wide-JAV | -0.0017 | -3.38 | 0.027 | -0.0009 | -0.21 | 0.027 | -0.0001 | -0.02 | 0.014 |
| | FCS-2L-wide-passive | -0.0017 | -3.35 | 0.027 | -0.0012 | -0.27 | 0.027 | -0.0002 | -0.03 | 0.014 |
| | SMC-JM-2L-DI | -0.0017 | -3.41 | 0.026 | -0.0013 | -0.28 | 0.027 | -0.0005 | -0.10 | 0.014 |
| | SMC-SM-2L-DI | -0.0018 | -3.50 | 0.026 | -0.0013 | -0.29 | 0.027 | -0.0005 | -0.09 | 0.014 |
| | SMC-JM-3L | -0.0014 | -2.85 | 0.027 | -0.0013 | -0.30 | 0.027 | -0.0005 | -0.09 | 0.014 |

Table S12: Performance of the 9 multiple imputation (MI) methods for estimating the main effect ($\beta_1 = -0.024$) and the quadratic effect ($\beta_3 = -0.009$) in the analysis model 3 when data are missing at random with inflated dependencies (MAR-inflated)

| No of higher-level clusters | Method | Main effect (-0.024) | | | | | | Quadratic effect (-0.009) | | | | | |
|---|---|---|---|---|---|---|---|---|---|---|---|---|---|
| | | Average Estimate | Bias | Relative Bias (%) | Emp SE | Model SE | Coverage | Average Estimate | Bias | Relative Bias (%) | Emp SE | Model SE | Coverage |
| | JM-1L-DI-wide | -0.029 | -0.0045 | 18.9 | 0.0090 | 0.0093 | 92.9% | -0.007 | 0.0024 | -26.8 | 0.0026 | 0.0035 | 96.0% |
| | JM-1L-DI-wide-JAV | -0.023 | 0.0015 | -6.3 | 0.0109 | 0.0113 | 95.1% | -0.004 | 0.0047 | -52.7 | 0.0045 | 0.0046 | 83.3% |

| | Method | | | | | | | | | | | |
|---|---|---|---|---|---|---|---|---|---|---|---|---|
| 40 clusters | FCS-1L-DI-wide | -0.029 | -0.0045 | 18.9 | 0.0090 | 0.0093 | 92.8% | -0.007 | 0.0024 | -26.6 | 0.0026 | 0.0035 | 96.4% |
| | FCS-1L-DI-wide-passive | -0.029 | -0.0045 | 18.9 | 0.0090 | 0.0093 | 92.7% | -0.007 | 0.0024 | -26.9 | 0.0026 | 0.0035 | 95.8% |
| | JM-2L-wide-JAV | -0.023 | 0.0014 | -5.9 | 0.0109 | 0.0113 | 95.2% | -0.004 | 0.0046 | -51.4 | 0.0045 | 0.0046 | 82.8% |
| | FCS-2L-wide-passive | -0.029 | -0.0046 | 19.0 | 0.0090 | 0.0093 | 93.1% | -0.007 | 0.0025 | -27.7 | 0.0026 | 0.0035 | 95.5% |
| | SMC-JM-2L-DI | -0.025 | -0.0006 | 2.7 | 0.0092 | 0.0093 | 94.8% | -0.009 | -0.0001 | 0.9 | 0.0038 | 0.0036 | 94.2% |
| | SMC-SM-2L-DI | -0.024 | 0.0000 | -0.2 | 0.0090 | 0.0093 | 95.0% | -0.009 | -0.0001 | 1.6 | 0.0035 | 0.0035 | 95.1% |
| | SMC-JM-3L | -0.024 | 0.0003 | -1.3 | 0.0093 | 0.0095 | 95.3% | -0.009 | -0.0002 | 1.9 | 0.0036 | 0.0036 | 95.2% |
| 10 clusters | JM-1L-DI-wide | -0.029 | -0.0046 | 19.1 | 0.0093 | 0.0094 | 91.7% | -0.006 | 0.0025 | -27.8 | 0.0026 | 0.0036 | 95.5% |
| | JM-1L-DI-wide-JAV | -0.023 | 0.0014 | -6.0 | 0.0112 | 0.0114 | 95.5% | -0.004 | 0.0046 | -51.1 | 0.0046 | 0.0046 | 83.9% |
| | FCS-1L-DI-wide | -0.029 | -0.0046 | 19.3 | 0.0093 | 0.0093 | 90.8% | -0.006 | 0.0025 | -28.1 | 0.0026 | 0.0035 | 95.0% |
| | FCS-1L-DI-wide-passive | -0.029 | -0.0046 | 19.1 | 0.0093 | 0.0093 | 92.0% | -0.006 | 0.0026 | -28.5 | 0.0026 | 0.0035 | 95.3% |
| | JM-2L-wide-JAV | -0.023 | 0.0014 | -5.6 | 0.0112 | 0.0114 | 95.9% | -0.004 | 0.0046 | -51.0 | 0.0046 | 0.0046 | 83.1% |
| | FCS-2L-wide-passive | -0.029 | -0.0046 | 19.0 | 0.0093 | 0.0093 | 91.7% | -0.006 | 0.0026 | -28.4 | 0.0026 | 0.0035 | 94.5% |
| | SMC-JM-2L-DI | -0.024 | -0.0004 | 1.6 | 0.0096 | 0.0093 | 94.0% | -0.009 | -0.0001 | 1.4 | 0.0038 | 0.0036 | 94.4% |
| | SMC-SM-2L-DI | -0.024 | 0.0000 | -0.1 | 0.0093 | 0.0093 | 94.8% | -0.009 | -0.0001 | 1.1 | 0.0035 | 0.0035 | 94.6% |
| | SMC-JM-3L | -0.024 | 0.0004 | -1.6 | 0.0097 | 0.0096 | 94.2% | -0.009 | -0.0002 | 1.8 | 0.0037 | 0.0037 | 94.7% |

Table S13: Performance of the 9 multiple imputation (MI) methods in estimating the variance components (VC) at level 3, 2 and 1 in the analysis model 3 when data are missing at random with inflated dependencies (MAR-inflated)

| No of higher-level clusters | Method | Level 3 VC (0.05) | | | Level 2 VC (0.45) | | | Level 1 VC (0.5) | | |
|---|---|---|---|---|---|---|---|---|---|---|
| | | Bias | Relative Bias (%) | Emp SE | Bias | Relative Bias (%) | Emp SE | Bias | Relative Bias (%) | Emp SE |
| 40 clusters | JM-1L-DI-wide | 0.0000 | 0.06 | 0.016 | -0.0004 | -0.09 | 0.026 | 0.0003 | 0.06 | 0.015 |
| | JM-1L-DI-wide-JAV | 0.0001 | 0.15 | 0.016 | -0.0001 | -0.02 | 0.026 | 0.0007 | 0.13 | 0.015 |
| | FCS-1L-DI-wide | 0.0000 | 0.05 | 0.016 | -0.0004 | -0.09 | 0.026 | 0.0003 | 0.06 | 0.015 |
| | FCS-1L-DI-wide-passive | 0.0000 | 0.05 | 0.016 | -0.0004 | -0.09 | 0.026 | 0.0003 | 0.06 | 0.015 |
| | JM-2L-wide-JAV | 0.0000 | 0.05 | 0.016 | -0.0001 | -0.01 | 0.026 | 0.0006 | 0.13 | 0.015 |
| | FCS-2L-wide-passive | 0.0000 | 0.02 | 0.016 | -0.0004 | -0.08 | 0.026 | 0.0003 | 0.07 | 0.015 |
| | SMC-JM-2L-DI | 0.0000 | 0.05 | 0.016 | -0.0005 | -0.11 | 0.026 | 0.0000 | -0.01 | 0.014 |
| | SMC-SM-2L-DI | 0.0000 | -0.10 | 0.016 | -0.0004 | -0.10 | 0.026 | 0.0000 | 0.00 | 0.015 |
| | SMC-JM-3L | 0.0000 | 0.02 | 0.016 | -0.0005 | -0.12 | 0.026 | 0.0000 | 0.00 | 0.015 |
| 10 clusters | JM-1L-DI-wide | -0.0014 | -2.81 | 0.027 | -0.0012 | -0.27 | 0.027 | -0.0002 | -0.04 | 0.014 |
| | JM-1L-DI-wide-JAV | -0.0014 | -2.73 | 0.027 | -0.0009 | -0.20 | 0.027 | 0.0001 | 0.02 | 0.014 |
| | FCS-1L-DI-wide | -0.0014 | -2.80 | 0.027 | -0.0012 | -0.28 | 0.027 | -0.0002 | -0.04 | 0.014 |
| | FCS-1L-DI-wide-passive | -0.0014 | -2.83 | 0.027 | -0.0012 | -0.28 | 0.027 | -0.0002 | -0.04 | 0.014 |
| | JM-2L-wide-JAV | -0.0014 | -2.79 | 0.027 | -0.0009 | -0.20 | 0.027 | 0.0001 | 0.02 | 0.014 |
| | FCS-2L-wide-passive | -0.0014 | -2.83 | 0.027 | -0.0012 | -0.28 | 0.027 | -0.0002 | -0.04 | 0.014 |
| | SMC-JM-2L-DI | -0.0014 | -2.81 | 0.027 | -0.0013 | -0.29 | 0.027 | -0.0005 | -0.10 | 0.014 |
| | SMC-SM-2L-DI | -0.0015 | -2.91 | 0.027 | -0.0014 | -0.30 | 0.027 | -0.0005 | -0.09 | 0.014 |
| | SMC-JM-3L | -0.0014 | -2.84 | 0.027 | -0.0014 | -0.31 | 0.027 | -0.0005 | -0.10 | 0.014 |

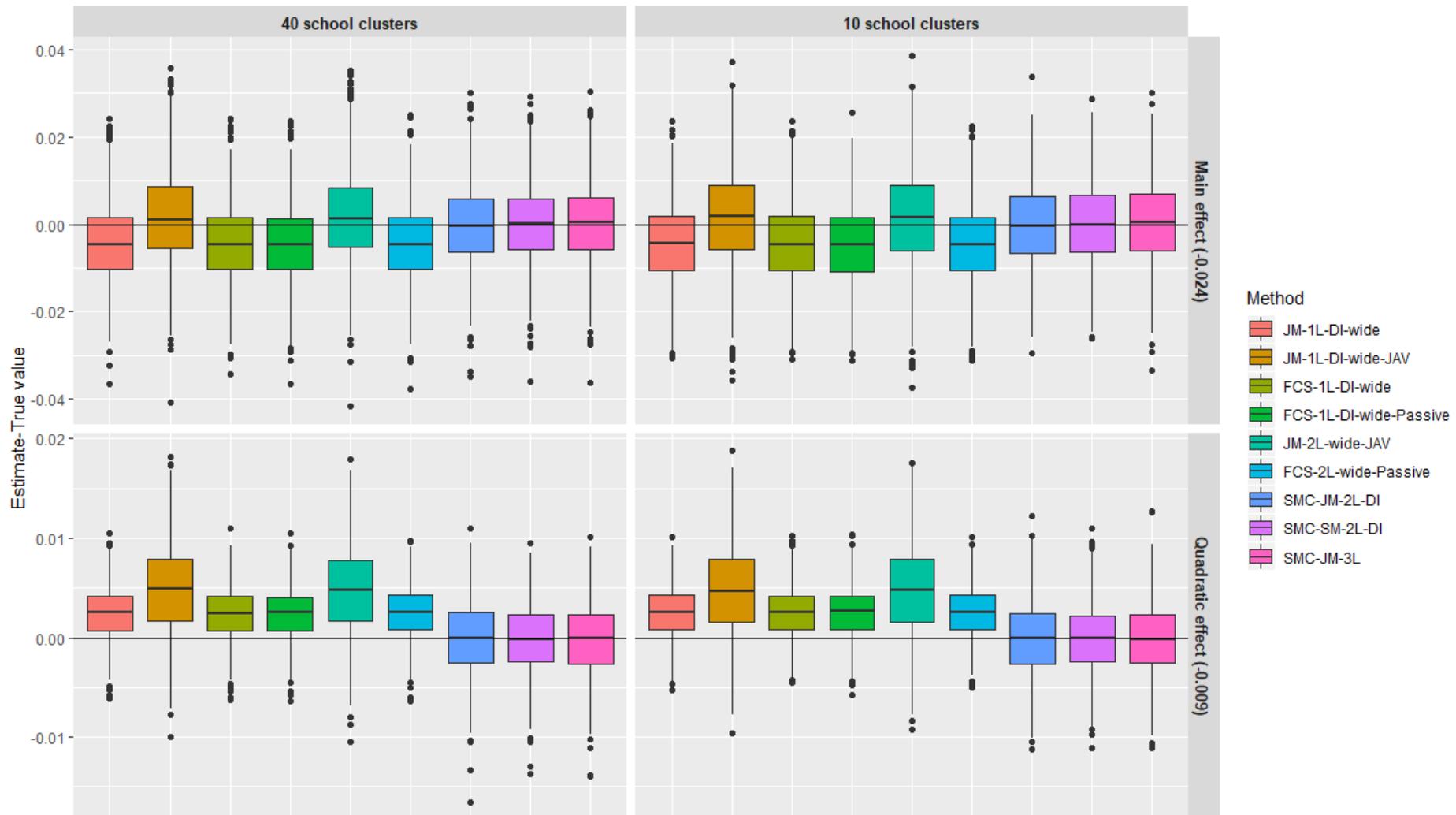

Figure S7: Distribution of the bias in the estimated regression coefficient for the main effect ($\beta_1$, $true\ value = -0.024$) and the quadratic effect ($\beta_3$, $true\ value = -0.009$) in analysis model 3 across the 1000 simulated datasets for the 9 multiple imputation (MI) approaches under two scenarios for number of higher level clusters( 40 school clusters and 10 school clusters) when data are missing at random with inflated dependencies (MAR-inflated)

The lower and upper margins of the boxes represent the 25$^{th}$ (Q$_1$) and the 75$^{th}$ (Q$_3$) percentiles of the distribution respectively. The whiskers extend to Q$_1$-1.5*(Q$_3$- Q$_1$) at the bottom and Q$_3$ +1.5*(Q$_3$- Q$_1$) at the top.

The following abbreviations are used to denote different MI methods, e.g., DI: dummy indicators, FCS: fully conditional specification, JM: joint modelling, SM: sequential modelling, SMC: substantive model compatible.

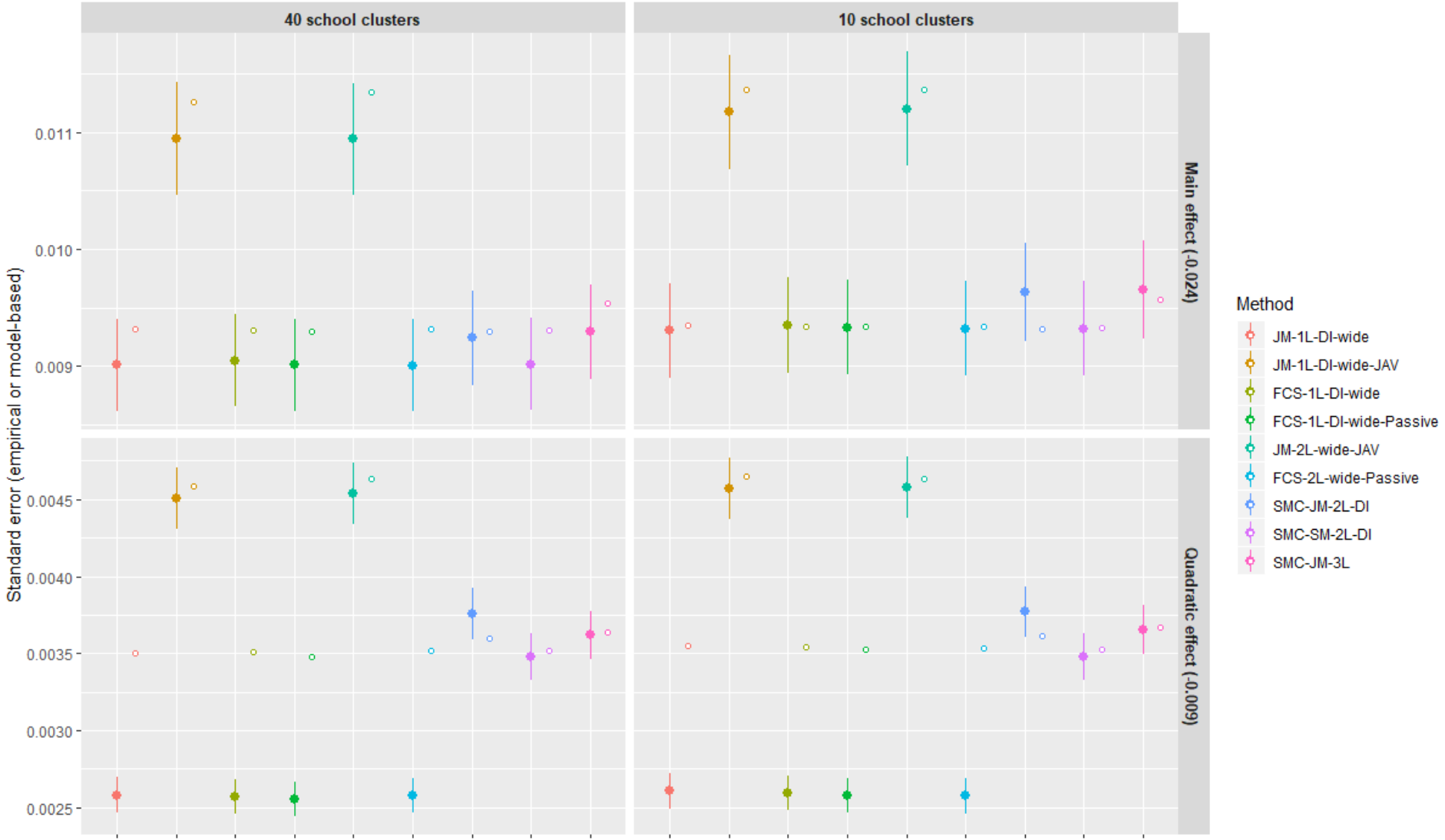

Figure S8: Empirical standard errors (filled circles with error bars showing ±1.96× Monte Carlo standard errors) and average model-based standard errors (hollow circles) in analysis model 3 from 1000 simulated datasets, for the 11 multiple imputation (MI) approaches under two scenarios for number of higher level clusters (40 school clusters and 10 school clusters) when data are missing at random with inflated dependencies (MAR-inflated)
The following abbreviations are used to denote different MI methods, e.g., DI: dummy indicators, FCS: fully conditional specification, JM: joint modelling, SM: sequential modelling, SMC: substantive model compatible.

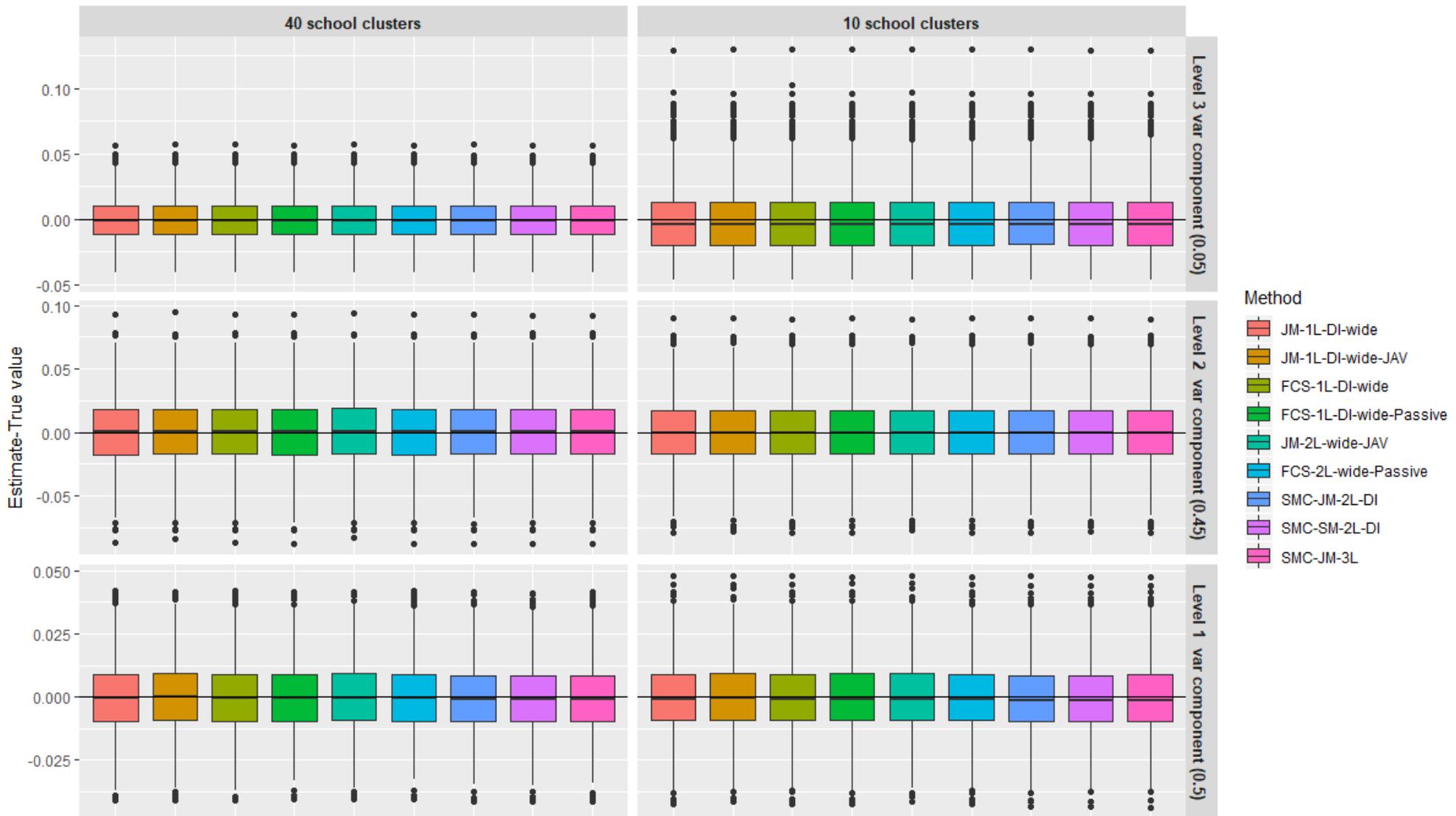

Figure S9: Distribution of the bias in the estimated variance components (VC) at level 1, 2 and 3 in analysis model 3 across the 1000 simulated datasets for the 9 multiple imputation (MI) approaches under two scenarios for number of higher level clusters (40 school clusters and 10 school clusters) when data are missing at random with inflated dependencies (MAR-inflated) The following abbreviations are used to denote different MI methods, e.g., DI: dummy indicators, FCS: fully conditional specification, JM: joint modelling, SM: sequential modelling, SMC: substantive model compatible.